\documentclass[preprint,showpacs]{revtex4}
\usepackage{graphicx}
\usepackage[bf,SL,BF]{subfigure}
\usepackage{psfrag}
\usepackage{color}
\usepackage{amssymb}
\usepackage{amsmath}
\usepackage{epstopdf}
\usepackage[normalem]{ulem}
\usepackage[T1]{fontenc}
\usepackage[utf8]{inputenc}
\usepackage{lmodern}
\usepackage{amsfonts}

\newcommand{\be}{\begin{eqnarray}}
\newcommand{\ee}{\end{eqnarray}}

 \newcommand{\bfk}{{\bf k}_{\perp}}

\newcommand{\bfb}{{\bf b}_{\perp}}

\newcommand{\bfp}{{\bf p}_{\perp}}

\newcommand{\pt}{p_{\perp}}
\newcommand{\bfPhp}{{\bf P}_{h\perp}}
\newcommand{\bfPh}{{\bf P}_{h}}
\newcommand{\Pht}{P_{h\perp}}

\newcommand{\avpsq}{\langle p_\perp^2\rangle}
\newcommand{\avksq}{\langle k_\perp^2 \rangle}
\newcommand{\avPhsq}{\langle P_{h\perp}^2\rangle}

\begin{document}
\title{ Model predictions for azimuthal spin asymmetries for HERMES and COMPASS kinematics }
\author{ Tanmay Maji}
\author{Dipankar Chakrabarti}
\affiliation{ Department of Physics, 
Indian Institute of Technology Kanpur,
Kanpur 208016, India}
\author{O.V. Teryaev}
\affiliation{Joint Institute of Nuclear Research, 141980 Dubna, Russia}
\date{\today}

\begin{abstract}
We have presented the results for the single and double spin asymmetries in semi-inclusive deep inelastic scatterings for proton 
in a light front quark-diquark model. The asymmetries generated by the T-even TMDs are discussed here. 
The model predictions are found to agree with the available data. We also present our model predictions 
for the Collins asymmetry for the future electron-ion collider experiments.
 
\end{abstract}
\pacs{14.20.Dh, 12.39.-x,12.38.Aw, 12.90.+b}
\maketitle
\section{Introduction\label{intro}}

Azimuthal spin asymmetries in semi-inclusive deep inelastic scattering(SIDIS) have been observed in many experiments. Measurements of 
azimuthal asymmetries are important to understand the transverse structure of the proton.
These asymmetries indicate existence of non-vanishing transverse momentum of interior quarks and collinear picture used for DIS  is not sufficient.
SIDIS cross section can be factorized into transverse momentum dependent parton distributions(TMDs) which contains the information
 of the distributions of quarks with transverse momentum in the parent proton   and the fragmentation functions(FFs) which describe the 
 hadronizations of the struck quarks into the detected hadrons.
 At leading twist, there are eight TMDs and two FFs for  unpolarized final hadrons. When the polarization of the final hadron is not detected, 
 the fragmentation is described by two FFs: chiral-even $D_1(z,k_\perp^2)$ which describe the fragmentation 
 of unpolarized hadron from a unpolarized quark, and chiral-odd 
$H_1^\perp(z,k_\perp^2)$ which is known as Collins function\cite{Collins:1992kk}, 
describes a 
 left-right asymmetry in the fragmentation of a transversely polarized quark($z$ is the energy fraction carried by the final hadron with 
 the transverse momentum $k_\perp$).  Out of the eight TMDs, Boer-Mulders and Sivers functions 
 $h_1^\perp(x,p_\perp^2)$ and $f_{1T}^\perp(x,p_\perp^2)$ are T-odd. To study the T-odd TMDs, one requires an  one gluon
 final state interaction which produces a complex phase in the wavefunctions. We do not consider the spin asymmetries caused by those TMDs here and 
 concentrate only on the azimuthal spin asymmetries involving T-even TMDs  in this article.
 
 Different single and double spin asymmetries observed in the angular distribution of the detected hadron, give crucial information about the TMDs.
 The TMD $h_{1L}^\perp(x,p_\perp^2)$ being chiral-odd can only be probed when it couples with  chiral-odd Collins function and is  accessed in 
 the single-spin-asymmetry(SSA) $A_{UL}$ with unpolarized lepton and longitudinally polarized target. Another chiral-odd TMD $h_1(x,p_\perp^2)$ 
 is accessed in SSA $A_{UT}$ requiring unpolarized lepton and transversely polarized target. The chiral-even TMD
 $g_{1T}^\perp(x,p_\perp^2)$ describes the probability of finding a longitudinal quark inside a transversely polarized proton and
 it can be obtained in double-spin-asymmetry(DSA) $A_{LT}$ involving longitudinally polarized lepton and transversely polarized proton.
 
Many phenomenological models have addressed the spin asymmetries. Most of the model calculations consider Gaussian ansatz for TMDs 
and FFs to extract the corresponding distribution functions from the fitting of the asymmetry data. A simultaneous extraction of 
Collins and transversity distribution is done by Anselmino $et. al.$\cite{Anselmino:2007fs,Anselmino:2008jk,Anselmino:2013vqa} from Collins asymmetry  data of
HERMES and COMPASS. Sivers function is extracted from Sivers asymmetry data in the Refs.\cite{Anselmino:2005nn, Anselmino:2005ea, Anselmino:2008sga}.

We calculate the Collins asymmetry as well as other single spin asymmetries where the  leading twist TMDs are calculated in 
light-front quark diquark model(LFQDM)\cite{Maji:2016yqo} and the fragmentation functions are taken from phenomenological 
parametrization \cite{Anselmino:2007fs,Anselmino:2013vqa,Kretzer:2001pz}. We have shown the Collins asymmetry for SIDIS process
$\ell N \to \ell^\prime X h$ at $\mu^2=2.5~GeV^2$ and compared with the experimental data of COMPASS and HERMES 
for $\pi^+$ and $\pi^-$ channels. 
One of the major challenges of comparing model results with experimental data is the scale evolution of TMDs. Till now, except the unpolarized TMDs,
the scale evolution of TMDs are not known. Since the model is defined at an initial scale, without proper scale evolution of the TMDs, the 
comparison of the model predictions with the data is incomplete.  Since the asymmetries are written as ratios of cross-sections, one may 
expect the scale evolution may get partially canceled in numerator and 
denominator and as a result the effect of evolution may not be very large.
In case of Collins asymmetry what we observe is that the effect of scale 
evolution gets partially canceled and  does not show much scale 
dependence but this is not true for all other  azimuthal asymmetries. For the asymmetries, 
  we keep the polarized TMDs at the initial scale and consider the scale evolution of the unpolarized TMD which is known. Thus the errors in the 
results are restricted in the polarized TMDs only. We also compare the results when the polarized TMDs are evolved in different 
approximation schemes. 
 Some  evolution ansatz may produce good agreements with the data for certain asymmetries but may fail in other cases. Unless, 
 we know the proper QCD evolution 
  of all the TMDs, it is not possible to favor one over the other.

A brief discussion on azimuthal asymmetries in SIDIS is given in Sec.\ref{sec_AA_SIDIS}. Model calculation of TMDs in light front quark model is 
discussed in the Sec.\ref{sec_model} followed by the TMDs evolution in brief. The model calculation of single spin asymmetry in SIDIS is
discussed in \ref{sec_SSAs} and  a comparison with experimental data of HERMES and COMPASS are also shown. The model prediction to the Double 
spin asymmetry data is presented in Sec.\ref{sec_DSAs}. 
\section{Azimuthal Asymmetries in SIDIS }\label{sec_AA_SIDIS}

\begin{figure}[htbp]
\includegraphics[width=7.2cm,clip]{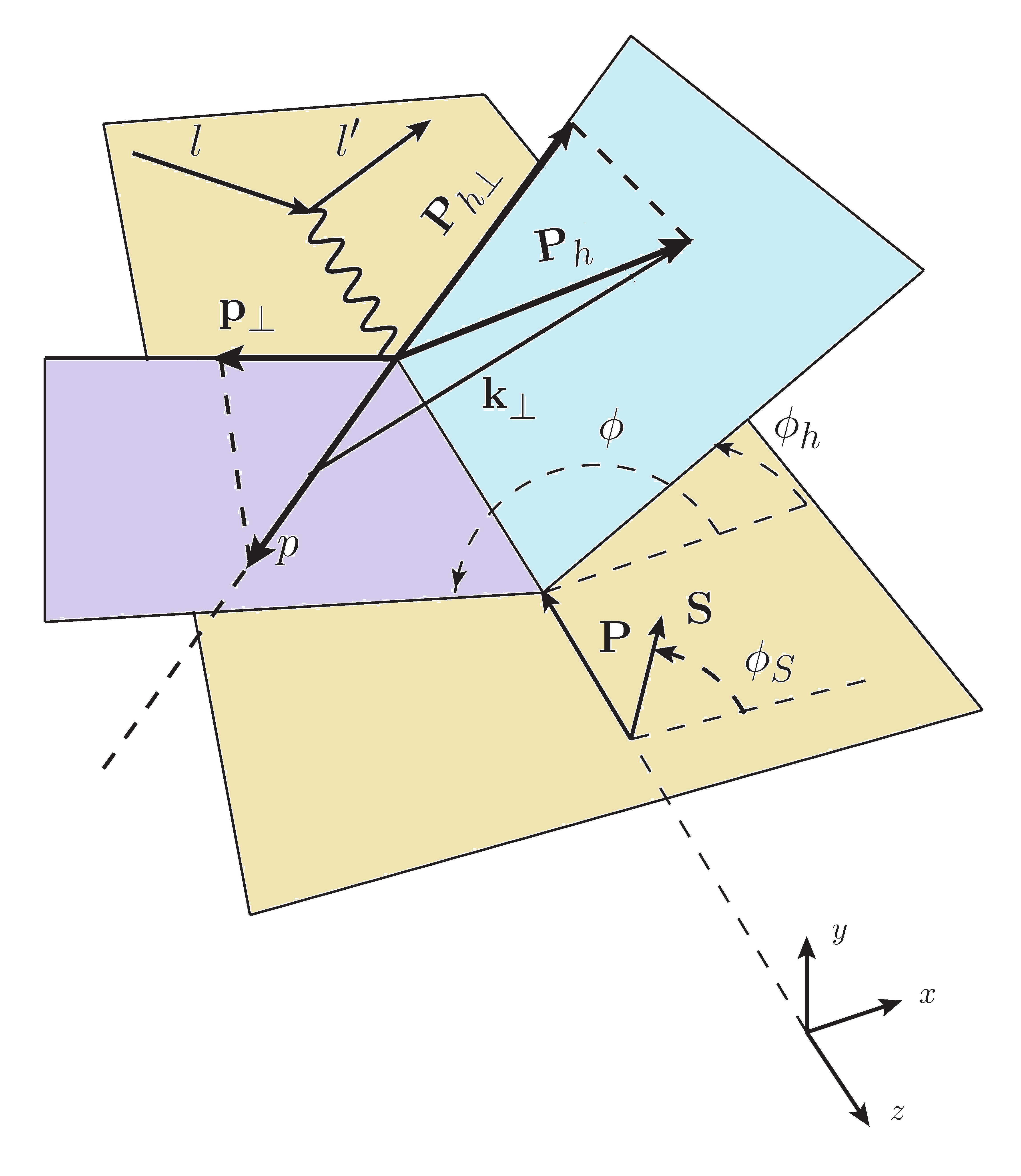}
\caption{\label{frame} $\gamma^*-P$ center of mass frame: produced hadron has a non-zero transverse momentum($\bfPhp$) in this frame and makes an azimuthal angle of $\phi_h$. The proton spin ($S$) has an azimuthal angle of $\phi_S$. All kinematics are given in text.}
\end{figure}

In the QCD factorization scheme the Semi-Inclusive Deep Inelastic 
Scattering(SIDIS) cross-section for the one photon exchange process
$\ell N \to \ell' h X$ is written as
\be 
d\sigma^{\ell N \to \ell' h X}=\sum_\nu \hat{f}_{\nu/P}(x,\bfp;Q^2)\otimes d\hat{\sigma}^{\ell q \to \ell q} \otimes \hat{D}_{h/\nu}(z,\bfk;Q^2);
\ee  
where the first term represents the transverse momentum dependent parton distribution functions(TMDs) which provides the probability of having a 
struck quarks of a particular polarization in a nucleon,  the second term represents the 
hard scattering which is a point like QED scattering mediated by a virtual photon and the third term is for fragmentation functions(FFs) 
which gives information about hadronizations fragmented from a quark. Such a scheme holds in small $\bfPhp$ and large $Q$ region, 
$P_{h\perp}^2 \simeq \Lambda^2_{QCD} \ll Q^2 $. At large $\bfPhp$ quark-gluon corrections and higher order pQCD corrections become 
important\cite{Bacchetta:2008af, Ji:2006br, Anselmino:2006rv}. 
The TMD factorization theorem is not proven generically for all the 
process. However, a 
 proof of the TMD factorization is presented for the 
SIDIS and the DY processes in \cite{Ji:2004wu,Ji:2004xq} and latter on used in 
\cite{Aybat:2011zv,Aybat:2011ge,Aybat:2011ta,Anselmino:2012aa}.
The kinematics of SIDIS is given in Fig.\ref{frame}. In the $\gamma^*-N$ center of mass frame, 
the kinematic variables are defined  as 
\be 
x=\frac{Q^2}{2(P.q)}=x_B \hspace{1.5cm}
y=\frac{P.q}{P.\ell}=\frac{Q^2}{s x} \hspace{1.5cm}
z=\frac{P.P_h}{P.q}=z_h. 
\ee 
In this frame, struck quark and diquark have equal and opposite transverse momentum and produced hadron gets a non-zero transverse momentum. 
Thus, momentum of the incoming proton $P \equiv (P^+, \frac{M^2}{P^+}, \textbf{0}_\perp)$ and of the virtual photon
$q \equiv (x_B P^+,\frac{Q^2}{x_B P^+}, \textbf{0}_\perp)$. Where $x_B=\frac{Q^2}{2P.q}$ is the Bjorken scaling with $Q^2=-q^2$. 
The struck quark of momentum $p \equiv (xP^+, \frac{p^2+|\bfp|^2}{xP^+}, \bfp)$ interact with the virtual photon and the diquark carries a 
momentum $p_D \equiv ((1-x)P^+, \frac{p^2+|\bfp|^2}{(1-x)P^+}, -\bfp)$. The produced hadron carries a momentum $\bfPh \equiv (P^+, P^-, \bfPhp)$. 
We use the light-cone convention $x^\pm = x^0 \pm x^3 $. The fractional energy transferred by the photon in the lab system is $y$ and the
energy fraction carried by the produced hadron is $z=\bfPh^-/k^-$. In this frame, though the incoming proton dose not have transverse momentum, 
the constituent quarks can have  non-zero transverse momenta which sum up to zero. $\bfp, \bfk$ and $\bfPhp$ are the transverse momentum 
carried by struck quark, fragmenting quark and fragmented hadron respectively. The relation between them, at $\mathcal{O}(\bfp/Q)$, is given by 
\be 
\bfk=\bfPhp-z\bfp
\ee
Here we consider one photon interaction only. The transverse momentum of produced hadron makes an azimuthal angle $\phi_h$ with respect to the 
lepton plane and transverse spin($S_P$) of the proton has an azimuthal angle $\phi_S$.

In the general helicity decomposition, the polarized SIDIS cross-section is 
written in terms of structure functions, at kinematic 
order $\bfp/Q$, as\cite{Anselmino:2011ch}
\be 
\frac{d\sigma^{\ell(S_\ell)+P(S_P)\to \ell' P_h X}}{dx_B dy dz d^2\bfPhp d\phi_S} &=& \frac{2\alpha^2}{s x y^2}\bigg\{\frac{1+(1-y)^2}{2}F_{UU}+(2-y)\sqrt{1-y}\cos\phi_h F^{\cos\phi_h}_{UU}+(1-y)\cos2\phi_h F^{\cos2\phi_h}_{UU}\nonumber\\
&+& S^L_P \bigg[(1-y)\sin2\phi_h F^{\sin2\phi_h}_{UL}+ (2-y)\sqrt{1-y}\sin\phi_h F^{\sin\phi_h}_{UL}\bigg] \nonumber\\
&+& S^L_P S^z_\ell \bigg[\frac{1-(1-y)^2}{2}F_{LL}+y\sqrt{1-y}\cos\phi_h F^{\cos\phi_h}_{LL}\bigg] \nonumber\\
&+&S^T_P\bigg[\frac{1+(1-y)^2}{2}\sin(\phi_h-\phi_S)F^{\sin(\phi_h-\phi_S)}_{UT}\nonumber\\
&+&(1-y)\bigg(\sin(\phi_h+\phi_S)F^{\sin(\phi_h+\phi_S)}_{UT}+\sin(3\phi_h-\phi_S)F^{\sin(3\phi_h-\phi_S)}_{UT}\bigg)\nonumber\\
&+&(2-y)\sqrt{(1-y)}\bigg(\sin\phi_S F^{\sin\phi_S}_{UT}+\sin(2\phi_h-\phi_S)F^{\sin(2\phi_h-\phi_S)}_{UT}\bigg)\bigg] \nonumber\\
&+& S^T_P S^z_\ell \bigg[\frac{1-(1-y)^2}{2}\cos(\phi_h-\phi_S)F^{\cos(\phi_h-\phi_S)}_{LT} \nonumber\\
&+& y\sqrt{1-y}\bigg(\cos\phi_S F^{\cos\phi_S}_{LT} + \cos(2\phi_h-\phi_S)F^{\cos(2\phi_h-\phi_S)}_{LT}\bigg)\bigg]\bigg\} \label{Cross}
\ee 
Where $S_\ell$ is the lepton polarization and $S^{L/T}_P$ represent the  
polarization of proton with longitudinally polarization(L) and transverse 
polarization(T) index at the superscript. The first three terms(first line) 
contribute 
to the unpolarized cross-section and the other terms contribute for different 
proton polarizations.  

The weighted structure functions, $F^{\mathcal{W}(\phi_h,\phi_S)}_{S_\ell S}$, are defined as 
\be 
F^{\mathcal{W}(\phi_h,\phi_S)}_{S_\ell S}&=&\mathcal{C}[\mathcal{W} \hat{f}(x,\bfp) \hat{D}(z,\bfk)] \nonumber\\
&=& \sum_\nu e^2_\nu  \int d^2\bfp d^2\bfk \delta^{(2)}(\bfPhp-z\bfp-\bfk) \mathcal{W}(\bfp,\bfPhp) \hat{f}^\nu(x,\bfp)\hat{D}^\nu(z,\bfk),\label{conv}
\ee
where  $\hat{f}^\nu(x,\bfp)$ and $\hat{D}^\nu(z,\bfk)$ represent leading twist TMDs and FFs respectively.
The above convolution integral is solved assuming Gaussian ansatz for TMDs in several models as well as in phenomenological extractions \cite{Boffi:2009sh, Anselmino:2013vqa, Anselmino:2007fs}.

The weighted structure functions contributing to SSAs are written in terms of convolutions of TMDs and FFs as\cite{Anselmino:2011ch}
\be 
F_{UU}&=&\mathcal{C}[ f^\nu_1 D^{h/\nu}_1],\label{SF_FUU}\\
\sin2\phi_h F^{\sin2\phi_h}_{UL}&=&\sin2\phi_h\mathcal{C}\bigg[\frac{(\bfPhp.\bfp)-2z(\hat{\bf{P}}_{h\perp}.\bfp)^2+
zp^2_\perp}{z M_h M}h^{\perp\nu}_{1L} H^{\perp\nu}_1\bigg], \label{SF_2h} \\
\sin\phi_h F^{\sin\phi_h}_{UL}&=&\sin\phi_h(\frac{-2}{Q})\mathcal{C}\bigg[\frac{p^2_\perp(P_{h\perp}-z\hat{\bf{P}}_{h\perp}.\bfp)}
{z M_h M}h^{\perp\nu}_{1L} H^{\perp\nu}_1\bigg], \label{SF_h}\\
\sin(\phi_h+\phi_S) F^{\sin(\phi_h+\phi_S)}_{UT}&=&\sin(\phi_h+\phi_S)\mathcal{C}\bigg[\frac{P_{h\perp}-z (\hat{\bf{P}}_{h\perp}.\bfp)}
{zM_h} h^{\nu}_{1} H^{\perp\nu}_1\bigg], \label{SF_Coll}\\
\sin(3\phi_h-\phi_S) F^{\sin(3\phi_h-\phi_S)}_{UT}&=&\sin(3\phi_h-\phi_S)\mathcal{C}\bigg[p^2_\perp\bigg(\frac{-P_{h\perp}+2P_{h\perp}
(\hat{\bf{P}}_{h\perp}.\hat{\bf{p}}_\perp)^2}{2zM_h M^2} \nonumber\\
&&-\frac{z p_\perp[4(\hat{\bf{P}}_{h\perp}.\hat{\bf{p}}_\perp)^3+3(\hat{\bf{P}}_{h\perp}.\hat{\bf{p}}_\perp)]}{2 z M_h M^2}\bigg) 
h^{\perp\nu}_{1T} H^{\perp\nu}_1\bigg],\label{SF_3hms}
\ee 
and the structure functions contributing to the DSAs are given by
\be 
F_{LL}&=&\mathcal{C}[ g^\nu_{1L} D^{h/\nu}_1],\label{SF_FLL}\\
\cos\phi_h F^{\cos\phi_h}_{LL}&=&\cos\phi_h \bigg(-\frac{2}{Q}\bigg)\mathcal{C}\bigg[(\hat{\textbf{P}}_{h\perp}.\bfp) g^\nu_{1L} D^{h/\nu}_1\bigg],\label{SF_FLL_cosh}\\
\cos(\phi_h-\phi_S) F^{\cos(\phi_h-\phi_S)}_{LT}&=&\cos(\phi_h-\phi_S) \mathcal{C}\bigg[\frac{(\hat{\textbf{P}}_{h\perp}.\bfp)}{M} g^\nu_{1T} D^{h/\nu}_1\bigg],\label{SF_FLT_coshms}\\
\cos\phi_S F^{\cos\phi_S}_{LT}&=&\cos\phi_S \bigg(-\frac{1}{Q}\bigg)\mathcal{C}\bigg[\frac{\pt^2}{M} g^\nu_{1T} D^{h/\nu}_1\bigg],\label{SF_FLT_cosS}\\
\cos(2\phi_h-\phi_S) F^{\cos(2\phi_h-\phi_S)}_{LT}&=&\cos(2\phi_h-\phi_S) \frac{1}{Q}\mathcal{C}\bigg[\frac{(\pt^2-2(\hat{\textbf{P}}_{h\perp}.\bfp)^2)}{M} g^\nu_{1T} D^{h/\nu}_1\bigg],\label{SF_FLT_coshms}
\ee
Where, $\mathcal{C}$ stands for the convolution as defined in Eq.(\ref{conv}) and  $f_1,h^\perp_{1L}, h_{1},h^\perp_{1T}, g_{1L}$ and $g_{1T}$ are 
the leading twist T-even TMDs which are functions of $x$ and $\bfp$. 
$D^{h/\nu}_1 \equiv D^{h/\nu}_1(z,\bfk)$ is the unpolarized FF and $H^\perp_1 
\equiv H^\perp_1(z,\bfk)$ is 
the Collins fragmentation function.
The contribution of above structure functions to the azimuthal spin asymmetries are discussed in the following sections.

In the SIDIS process, asymmetry is observed experimentally during the measurement of angular distribution of produced hadrons. The azimuthal 
asymmetries in SIDIS process are defined as
\be 
A_{S_\ell S_P}&=&\frac{d\sigma^{\ell(S_\ell) P(S_P) \to \ell' h X}-d\sigma^{\ell(S_\ell) P(-S_P)\to \ell' h X}}{d\sigma^{\ell(S_\ell) P(S_P) 
\to \ell' h X}+d\sigma^{\ell(S_\ell) P(-S_P) \to \ell' h X}}. 
\ee
Note that, $ d\sigma^{\ell(S_\ell) P(S_P) \to \ell' h X} $ is a short hand notation 
of $\frac{d\sigma^{\ell(S_\ell)+P(S)\to \ell' P_h X}}{dx_B dy dz d^2\bfPhp d\phi_S}$ of Eq.(\ref{Cross}). Thus using Eq.(\ref{Cross}) 
the asymmetries can be expressed in terms of structure functions and then as a convolution of leading twist TMDs and FFs. Since, 
in the cross-section, each structure functions comes with a defined angular coefficient, the contribution of single TMDs can be extracted by
introducing corresponding weight factor(and integrating over $\phi_h$ and $\phi_S$) in the definition of azimuthal asymmetry as
\be 
A^{\mathcal{W}(\phi_h,\phi_S)}_{S_\ell S_P}&=&2\frac{\int d\phi_h d\phi_S [d\sigma^{\ell(S_\ell) P(S_P) \to \ell' h X}-d\sigma^{\ell(S_\ell) 
P(-S_P) \to \ell' h X}]\mathcal{W}(\phi_h,\phi_S)}{\int d\phi_h d\phi_S [d\sigma^{\ell(S_\ell) P(S_P) \to \ell' h X}+d\sigma^{\ell(S_\ell) P(-S_P) \to \ell' h X}]} \label{Asy_W}
\ee
Where the function $\mathcal{W}(\phi_h,\phi_S)$ is a weight factor which project out corresponding asymmetry. For example, Collins asymmetry 
can be extracted by the weight factor $\mathcal{W}(\phi_h,\phi_S)=\sin(\phi_h+\phi_S)$ for a transversely polarized proton interacting with a 
unpolarized lepton beam. There are many more weighed asymmetries in SIDIS 
process some of them are observed experimentally. Here we will 
restrict ourselves to the asymmetries which has contribution from T-even leading twist TMDs and fragmentations. A detailed calculation of 
the different SSAs and DSAs are discussed in sec.\ref{sec_SSAs} and \ref{sec_DSAs}.

\section{Model Calculations}\label{sec_model}
Before we get into the asymmetries, let us discuss about the model in brief. Since different asymmetries have contribution from 
different leading twist TMDs and FFs, we give model prediction to the azimuthal spin asymmetries measured by HERMES, COMPASS experiments 
by calculating the leading twist TMDs in a recently proposed  light-front quark-diquark model(LFQDM)\cite{Maji:2016yqo}. In this model,
the wave functions are constructed in the framework of soft-wall AdS/QCD prediction. As we mentioned before, we concentrate on the 
asymmetries related to the T-even TMDs at the leading twist. The FFs $D^{h/\nu}_1(z,\bfk^2)$ and $H^{\perp\nu}_1(z,\bfk^2)$ are taken as 
a phenomenological input from Refs.\cite{Anselmino:2007fs, Anselmino:2013vqa, Kretzer:2001pz}. The model calculation(LFQDM) of TMDs are discussed briefly in the following subsection. 
\subsection{TMDs in LFQDM}
In this subsection we briefly discuss about calculation of leading twist T-even TMDs in the recently proposed model LFQDM\cite{Maji:2016yqo}. 
In this model,the proton state is written as two particle bound state of a quark 
and a diquark having a spin-flavor $SU(4)$ structure.
 \be 
|P; \pm\rangle = C_S|u~ S^0\rangle^\pm + C_V|u~ A^0\rangle^\pm + C_{VV}|d~ A^1\rangle^\pm. \label{PS_state}
\ee 
Where $\mid u~ S^0\rangle,~|u~ A^0\rangle$ and $|d~ A^1\rangle$ are two particle states having isoscalar-scalar, isoscalar-axialvector 
and isovector-axialvector diquark respectively\cite{Jakob:1997wg,Bacchetta:2008af}. The states are written in two particle Fock 
state expansion with $J^z =\pm1/2$ for both the scalar and the axial-vector diquarks\cite{Maji:2016yqo}. The two particle Fock state wave 
functions are adopted from soft-wall AdS/QCD prediction \cite{Brodsky:2007hb, deTeramond:2012rt} and modified as
\be
\varphi_i^{(\nu)}(x,\bfp)=\frac{4\pi}{\kappa}\sqrt{\frac{\log(1/x)}{1-x}}x^{a_i^\nu}(1-x)^{b_i^\nu}\exp\bigg[-\delta^\nu\frac{\bfp^2}{2\kappa^2}\frac{\log(1/x)}{(1-x)^2}\bigg].
\label{LFWF_phi}
\ee
 We use the AdS/QCD scale parameter $\kappa =0.4$ GeV as determined in \cite{Chakrabarti:2013gra}. The parameters $a_i^\nu,b_i^\nu$ and $\delta^\nu$ are fixed by fitting the Dirac and Pauli form factors. The quarks are  assumed  to be  massless. 

In the light-front formalism, the TMDs correlator at equal light-front time $z^+=0$  is defined for SIDIS as
\be
\Phi^{\nu [\Gamma]}(x,\textbf{p}_{\perp};S)&=&\frac{1}{2}\int \frac{dz^- d^2z_T}{2(2\pi)^3} e^{ip.z} \langle P; S|\overline{\psi}^\nu (0)\Gamma \mathcal{W}_{[0,z]} \psi^\nu (z) |P;S\rangle\Bigg|_{z^+=0} \label{TMD_cor}
\ee
 for different Dirac structures $\Gamma=\gamma^+,\gamma^+\gamma^5$ and $i\sigma^{j+}\gamma^5$. Where $x ~(x=p^+/P^+)$ is the longitudinal momentum fraction carried by the struck quark of helicity $\lambda$. 
 The proton spin components are $S^+ = \lambda_N \frac{P^+}{M},~ S^- = \lambda_N\frac{P^-}{M},$ and $ S_T $ with helicity $\lambda_N$.
In the leading twist, the TMD correlator is connected with the corresponding TMDs for different Dirac structures as\\
\begin{eqnarray}
\Phi^{\nu [\gamma^+]}(x,\textbf{p}_{\perp};S)&=& f_1^\nu (x,\textbf{p}_{\perp}^2) - \frac{\epsilon^{ij}_Tp^i_\perp S^j_T}{M}f^{\perp  \nu} _{1T}(x,\textbf{p}_{\perp}^2),\label{Phi_1}\\
\Phi^{\nu [\gamma^+ \gamma^5]}(x,\textbf{p}_{\perp};S) &=&  \lambda g_{1L}^\nu (x,\textbf{p}_{\perp}^2) + \frac{\textbf{p}_{\perp}.\textbf{S}_T}{M} g^\nu _{1T}(x,\textbf{p}_{\perp}^2),\label{Phi_2}\\
\Phi^{\nu [i \sigma^{j +}\gamma^5]}(x,\textbf{p}_{\perp};S)& = & S^j_T h_1^\nu (x,\textbf{p}_{\perp}^2) + \lambda\frac{p^j_\perp}{M}h^{\perp  \nu} _{1L}(x,\textbf{p}_{\perp}^2)\nonumber\\
&&+ \frac{2 p^j_\perp \textbf{p}_{\perp}.\textbf{S}_T - S^j_T \textbf{p}^2_{\perp}}{2M^2} h^{\perp  \nu} _{1T}(x,\textbf{p}_{\perp}^2) - \frac{\epsilon_T^{ij}p^i_{\perp}}{M}h^{\perp  \nu} _1(x,\textbf{p}_{\perp}^2).\label{Phi_3}
\end{eqnarray}
The transversity TMD $h^{\nu }_1 (x,\bfp)$ is given as 
\begin{eqnarray}
h^{\nu }_1 (x,\textbf{p}_{\perp}^2) &=& h^{\nu }_{1T} (x,\textbf{p}_{\perp}^2) + \frac{\bfp^2}{2M^2}h^{\perp  \nu} _{1T} (x,\textbf{p}_{\perp}^2).\label{h1q}
\end{eqnarray} 
The T-odd TMDs $f^\perp_{1T}$ and $h^\perp_1$ vanish as no gluon degrees of freedom is considered here. The one gluon final state interaction is needed to calculate the T-odd TMDs. The final state interaction generates a phase term in the wave functions which give rise to a non vanishing T-odd TMDs\cite{Brodsky:2002cx}.

In this model, a explicit form of the wave functions is given in \cite{Maji:2016yqo}. 
Using those in the correlator of Eq.(\ref{TMD_cor}) and comparing with the 
decompositions of Eq.(\ref{Phi_1}-\ref{Phi_3}), the leading twist T-even TMDs 
contributing to the SSA reads explicitly as\cite{Maji:2017bcz}
\be 
{f}^{\nu  }_1(x,\bfp^2)&=&\bigg(C^2_SN^{\nu 2}_S+C^2_A\big(\frac{1}{3}N^{\nu 2}_0+\frac{2}{3}N^{\nu 2}_1\big)\bigg)\frac{\ln(1/x)}{\pi\kappa^2}\bigg[T^\nu_1(x) +\frac{\bfp^2}{M^2} T^\nu_2(x)\bigg]\exp\big[-R^\nu(x)\bfp^2\big], \nonumber\\ 
\label{TMD_f1}
{h}^{\nu  }_1(x,\bfp^2) &=& \bigg(C^2_SN^{\nu 2}_S-C^2_A\frac{1}{3}N^{\nu 2}_0\bigg) \frac{\ln(1/x)}{\pi\kappa^2}T^\nu_1(x)\exp\big[-R^\nu(x)\bfp^2\big], \label{TMD_h1}\\
{h}^{\nu\perp}_{1L}(x,\textbf{p}^2_{\perp})&=& -\bigg(C^2_SN^{\nu 2}_S+C^2_A\big(\frac{1}{3}N^{\nu 2}_0 - \frac{2}{3}N^{\nu 2}_1\big)\bigg)\frac{2\ln(1/x)}{\pi\kappa^2} T^\nu_3(x)\exp\big[-R^\nu(x)\bfp^2\big],\label{TMD_h1Lp}\\
{h}^{\nu\perp}_{1T}(x,\textbf{p}^2_{\perp})&=& - \bigg(C^2_SN^{\nu 2}_S-C^2_A\frac{1}{3}N^{\nu 2}_0\bigg) \frac{2\ln(1/x)}{\pi\kappa^2}T^\nu_2(x)\exp\big[-R^\nu(x)\bfp^2\big].\label{TMD_h1Tp}
\ee
The T-even TMDs contributing to the DSAs are given by 
\be 
g^{\nu}_{1L}(x,\bfp^2)&=&\bigg(C^2_SN^{\nu 2}_S+C^2_A\big(\frac{1}{3}N^{\nu 2}_0-\frac{2}{3}N^{\nu 2}_1\big)\bigg)\frac{\ln(1/x)}{\pi\kappa^2}\bigg[T^\nu_1(x) -\frac{\bfp^2}{M^2} T^\nu_2(x)\bigg]\exp\big[-R^\nu(x)\bfp^2\big],\nonumber \label{TMD_g1} \\
g^{\nu}_{1T}(x,\textbf{p}^2_{\perp})&=& \bigg(C^2_SN^{\nu 2}_S-C^2_A\frac{1}{3}N^{\nu 2}_0\bigg)\frac{2\ln(1/x)}{\pi\kappa^2} T^\nu_3(x)\exp\big[-R^\nu(x)\bfp^2\big].\label{TMD_g1T}\\
\ee
Where 
\be
T^\nu_1(x)&=& x^{2a^{\nu}_1}(1-x)^{2b^{\nu}_1-1}, \nonumber\\
T^\nu_2(x)&=& x^{2a^{\nu}_2 -2}(1-x)^{2b^{\nu}_2-1},\label{Fx}\\
T^\nu_3(x)&=& x^{a^{\nu}_1 +a^{\nu}_2-1}(1-x)^{b^{\nu}_1+b^{\nu}_2-1},\nonumber\\
R^\nu(x)&=&\delta^\nu \frac{\ln(1/x)}{\kappa^2 (1-x)^2}.\nonumber
\ee
The values of the model parameters $a^\nu_i, b^\nu_i(i=1,2)$ and $\delta^\nu$ are given in \cite{Maji:2016yqo} at initial scale $\mu_0=0.8~GeV$ with the AdS/QCD scale parameter $\kappa=0.4$ GeV \cite{Chakrabarti:2013gra}. The pre-factors containing $C_j(j=S,V,VV)$ and $N_k(j,k=S,0,1)$ are the normalized constants which satisfy the quark counting rules for unpolarized TMDs. 
The subscript $A$ represents $V$ and $VV$ for $u$ and $d$ quarks respectively. 
Note that the normalization constant $N^d_S=0$ for $d$ quarks.

\subsection{Fragmentation functions }
We use Gaussian ansatz for fragmentations functions as discussed in Ref.\cite{Anselmino:2007fs, Anselmino:2013vqa}. 
\be 
D^{h/\nu}_1(z,\bfk)&=&D^{h/\nu}_1(z)\frac{e^{-\bfk^2/\avksq}}{\pi \avksq}\label{FF_D1}\\
\frac{2 k_\perp}{z M_h} H^{\perp\nu}_1(z,\bfk)&=&2\mathcal{N}^C_\nu(z) D^{h/\nu}_1(z)h(k_\perp)\frac{e^{-\bfk^2/\avksq}}{\pi \avksq} \label{FF_H1}
\ee
with 
\be 
\mathcal{N}^C_\nu(z)&=&N^C_\nu z^{\rho_1} (1-z)^{\rho_2}\frac{(\rho_1+\rho_2)^{(\rho_1+\rho_2)}}{\rho^{\rho_1}_1\rho^{\rho_2}_2}\\
h(k_\perp)&=&\sqrt{2 e} \frac{k_\perp}{M_h}e^{-\bfk^2/M^2_h}
\ee
Where the hadron of momentum $\bfPh$ and of energy fraction $z=P_h^-/k^-$ is 
produced from a fragmenting quark of momentum $\textbf{k}$. The values of the 
parameters are listed in \cite{Anselmino:2013vqa} and  $D^{h/\nu}_1(z)$ is taken 
from the phenomenological extraction\cite{Kretzer:2001pz}. The average value of 
the momentum is taken\cite{Anselmino:2012aa} as $\avksq= 0.2~GeV^2$. 

\subsection{TMD evolutions}\label{evol}
The $Q^2$ evolution of unpolarized TMD and unpolarized fragmentations functions are proposed in \cite{Aybat:2011zv}. 
An extension of the unpolarized TMD evolution is presented in \cite{Aybat:2011ge} and provides a framework to the scale evolution of 
spin-dependent distributions.  
The QCD evolution of TMDs in the coordinate space is defined\cite{Aybat:2011zv,Anselmino:2012aa} as
\be 
\tilde{F}(x,\bfb ;\mu)=\tilde{F}(x,\bfb ;\mu_0) \exp\bigg(\ln\frac{\mu}{\mu_0}\tilde{K}(b_\perp;\mu)+\int^\mu_{\mu_0}\frac{d\mu'}{\mu'}\gamma_F\big(\mu',\frac{\mu^2}{\mu'^2}\big)\bigg).\label{Ans_evol}
\ee
Where $\tilde{F}(x,\bfb ;\mu_0)$ is the TMDs at the initial scale $\mu_0$ and the exponential function contains the QCD evolution of the 
corresponding TMDs. The function $\tilde{K}(b_\perp;\mu)$ is given by\cite{Aybat:2011ge} 
\be 
\tilde{K}(b_\perp;\mu)=\tilde{K}(b_*;\mu_b)+\bigg[\int^{\mu_b}_\mu \frac{d\mu'}{\mu'}\gamma_K(\mu')\bigg]-g_K(b_T),\label{kt_bpmu}
\ee
where,
\be 
\tilde{K}(b_*;\mu_b)=-\frac{\alpha_s C_F}{\pi}[\ln(b_*^2\mu^2_b)-\ln(4)+2\gamma_E],\label{kt_bstmub}\\
b_*(b_T)=\frac{b_T}{\sqrt{1+\frac{b^2_T}{b^2_{max}}}} ~~~ ; ~~~ \mu_b=\frac{C_1}{b_*(b_T)} \label{bst_mub}
\ee
at $\mathcal{O}(\alpha_s)$\cite{Collins:1984kg,Collins:1984kg}.
We adopt a particular choice for the constant $C_1=2 e^{-\gamma_E}$\cite{Aybat:2011zv,Aybat:2011ge}, with the Euler constant $\gamma_E=0.577$ \cite{Collins:1984kg}. 
In the SIDIS, non-perturbative function $g_K(b_T)$ is parametrized \cite{Anselmino:2012aa, Aybat:2011ge, Landry:2002ix} as $g_K(b_T)=\frac{1}{2} g_2 b^2_T$ 
with $g_2=0.68~GeV^2$ and $b_{max}=0.5~GeV^{-1}$. This prescription overestimates the evolution for the Drell-Yan process as 
discussed in \cite{Sun:2013hua}.
\begin{figure}
\includegraphics[scale=0.4]{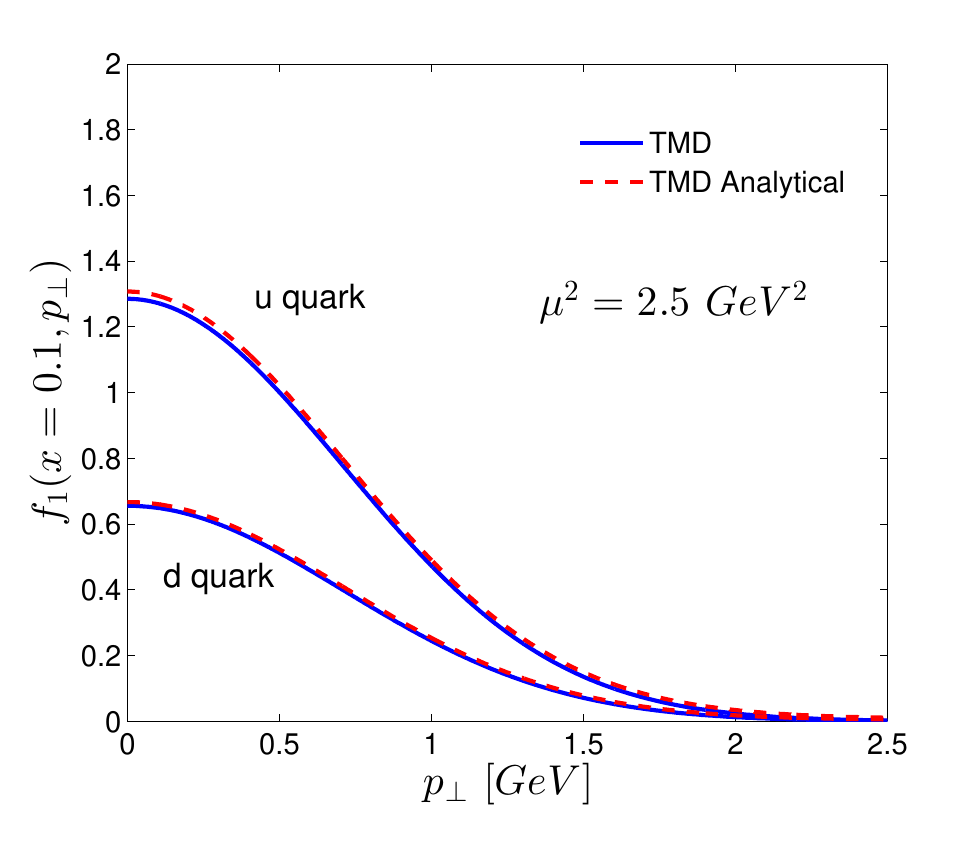}
\caption{\label{fig_TMD_analy} TMDs evolution: blue continuous lines represent QCD evolution with the kernel $\tilde{R}(\mu,\mu_0,b_T)$ and the red dashed lines represent the evolution with reduced kernel $R(\mu,\mu_0)$. }
\end{figure}
Using Eq.(\ref{kt_bpmu},\ref{kt_bstmub}, \ref{bst_mub}) the evolution Eq.(\ref{Ans_evol}) can be written as 
\be 
\tilde{F}(x,\bfb ;\mu)=\tilde{F}(x,\bfb ;\mu_0)\tilde{R}(\mu,\mu_0,b_T)\exp \bigg[-g_K(b_T)\ln(\frac{\mu}{\mu_0})\bigg],
\ee
with the kernel
\be  
\tilde{R}(\mu,\mu_0,b_T)=\exp\bigg[\ln\frac{\mu}{\mu_0}\int^{\mu_b}_\mu \frac{d\mu'}{\mu'}\gamma_K(\mu')+\int^\mu_{\mu_0}
\frac{d\mu'}{\mu'}\gamma_F\big(\mu',\frac{\mu^2}{\mu'^2}\big)\bigg].
\ee 
Here we consider the LO evolution. The anomalous dimensions are given by
\be  
\gamma_F\big(\mu',\frac{\mu^2}{\mu'^2}\big)&=&\alpha_s(\mu')\frac{C_F}{\pi}\bigg(\frac{3}{2}-\ln\frac{\mu^2}{\mu'^2}\bigg),\\
\gamma_K(\mu')&=&\alpha_s(\mu')\frac{C_F}{\pi}.
\ee
In the kernel $\tilde{R}(\mu,\mu_0,b_T)$, the impact parameter ($b_\perp$) dependency comes from the upper limit  $\mu_b$ in 
the  $\mu^\prime$ integration. 
This evolution equation can be solved analytically by making an approximation on $b_\perp$ as discussed in \cite{Anselmino:2012aa}. Eq.(\ref{bst_mub})indicates
that the $\mu_b$ converges to a constant value $\mu_b=C_1/b_{max}$ at the limit $b_\perp \to \infty $.  The large $b_\perp$ approximation ensures
a low $p_\perp$ region in this framework. Therefore under this approximation, the kernel $\tilde{R}(\mu,\mu_0,b_T)$ reduces to $R(\mu,\mu_0)$ and
the evolution equation can be integrated analytically. We compare the evolution of $f^\nu_1$ produced by the two kennels $\tilde{R}(\mu,\mu_0,b_T)$
and $R(\mu,\mu_0)$ at the scale $\mu^2=2.5~GeV^2$. We take a fixed value of $x=0.1$. We observed very insignificant difference in the QCD
evolution generated with $\tilde{R}(\mu,\mu_0,b_T)$ kernel  and with the reduced kernel $R(\mu,\mu_0)$ as shown in Fig.\ref{fig_TMD_analy}. 
Therefore, we calculate the SSAs at the scale $\mu^2=2.5~GeV^2$ by evolving the TMDs in reduced QCD evolution and compare with the
experimental data. Where the FFs are adopted from phenomenological parametrization at the scale $\mu^2=2.5~GeV^2$. 

\section{Single spin Asymmetries in LFQDM}\label{sec_SSAs}

The Single Spin Asymmetry(SSA) is measured when the target is polarized with respect to the beam direction. In the SIDIS processes,
the SSAs associated with unpolarized lepton(U) beam and transversely polarized proton(T) target is defined as 
\be 
A_{UT}&=&\frac{d\sigma^{\ell P^\uparrow \to \ell' h X}-d\sigma^{\ell P^\downarrow \to \ell' h X}}{d\sigma^{\ell P^\uparrow \to \ell' h X}+d\sigma^{\ell P^\downarrow \to \ell' h X}} \nonumber\\
&=&\frac{d\sigma^{\ell P(S_T) \to \ell' h X}-d\sigma^{\ell P(-S_T)\to \ell' h X}}{d\sigma^{\ell P(S_T) \to \ell' h X}+d\sigma^{\ell P(-S_T) \to \ell' h X}}. 
\ee
Where $\uparrow,\downarrow$ at the superscript of $P$ represent the transverse spin of the target proton.
From Eq.(\ref{Cross}) we can write the numerator of $A_{UT}$ as 
\be 
\frac{d\sigma^{\ell P^\uparrow \to \ell' h X}-d\sigma^{\ell P^\downarrow \to \ell' h X}}{dx_B dy dz d^2\bfPhp d\phi_S}&=& \frac{2\alpha^2}
{s x y^2}2\bigg[\frac{1+(1-y)^2}{2}\sin(\phi_h-\phi_S)F^{\sin(\phi_h-\phi_S)}_{UT}\nonumber\\
+&&\!\!\!\!\!\!\!\!(1-y)\bigg(\sin(\phi_h+\phi_S)F^{\sin(\phi_h+\phi_S)}_{UT}+\sin(3\phi_h-\phi_S)F^{\sin(3\phi_h-\phi_S)}_{UT}\bigg)\nonumber\\
+&&\!\!\!\!\!\!\!(2-y)\sqrt{(1-y)}\bigg(\sin\phi_S F^{\sin\phi_S}_{UT}+\sin(2\phi_h-\phi_S)F^{\sin(2\phi_h-\phi_S)}_{UT}\bigg)\bigg].\label{N_UT}
\ee
The first term  corresponds to the Sivers asymmetry which has contribution from 
Sivers functions($f^{\perp \nu}_{1T}$) and unpolarized FFs. 
The second term  corresponds to the Collins asymmetry which has contribution from transversity TMD ($h^\nu_1$) and Collins fragmentation 
function($H^{\perp h/\nu}_{1}$). The third term has contribution from pretzelocity distribution($h^{\perp\nu}_{1T}$). 
The fourth and fifth terms have contributions from multiple TMDs and FFs. Among these five SSAs, only two of them
$A^{\sin(\phi_h+\phi_S)}_{UT}(x,z,\bfPhp,y)$ and $A^{\sin(3\phi_h-\phi_S)}_{UT}(x,z,\bfPhp,y)$,  involve T-even TMDs and will be discussed here.

From Eq.(\ref{Cross}), the denominator can be written as 
\be 
\frac{d\sigma^{\ell P^\uparrow \to \ell' h X} + d\sigma^{\ell P^\downarrow \to \ell' h X}}{dx_B dy dz d^2\bfPhp d\phi_S}&=& \frac{2\alpha^2}{s x y^2}2 \bigg[\frac{1+(1-y)^2}{2}F_{UU}+(2-y)\sqrt{1-y}\cos\phi_h F^{\cos\phi_h}_{UU} \nonumber\\
&+&(1-y)cos2\phi_h F^{\cos2\phi_h}_{UU}\bigg]. \label{D_UT}
\ee
We extract the Collins asymmetry by introducing appropriate weighted factor $\sin(\phi_h+\phi_S)$ in Eq.(\ref{Asy_W}) and write in terms of
structure functions as
\be 
A^{\sin(\phi_h+\phi_S)}_{UT}(x,z,\bfPhp,y)&=& \frac{4\pi^2\alpha^2\frac{(1-y)}{s x y^2} F^{\sin(\phi_h+\phi_S)}_{UT}(x,z,\bfPhp,y)  }
{2\pi^2\alpha^2\frac{1+(1-y)^2}{s x y^2}F_{UU}(x,z,\bfPhp,y)}. \label{Coll_Asy}
\ee
The Collins asymmetry provides a correlation between the transverse 
polarization of the fragmenting  quark in a transversely polarized proton and 
the transverse momentum of the final hadron. 
 Since 
helicity is conserved in hard process, the chiral-odd TMD $h_1(x,\bfp)$ has to 
be convoluted with a chiral-odd FF, which is Collins function. Unlike Sivers 
function which differs by a sign for SIDIS and Drell-Yan processes, Collins 
function is same in both processes. 
In the SIDIS process we consider, a transversely polarized quark is scattered 
out of transversely polarized proton with the probability provided by 
transversity distribution $h_1(x,\bfp)$ and fragmented to a hadron with 
probability given by Collins function $H^\perp_1(z,\bfk)$. 
 The transverse polarization of the initial proton gets transferred to the 
final state by the hard scattering which produces an azimuthal spin asymmetry 
in the final hadron about the ``jet axis''. 

The azimuthal dependence in the structure function $F^{\sin(\phi_h+\phi_S)}_{UT}$,  given in Eq.(\ref{SF_Coll}), can be written in 
terms of $\phi$ as
\be 
\sin(\phi_h+\phi_S) F^{\sin(\phi_h+\phi_S)}_{UT}&=& \sum_\nu e^2_\nu \int d^2\bfp\frac{P_{h\perp}}{k_\perp} \bigg(\sin(\phi_h+\phi_S)-z\frac{p_\perp}{P_{h\perp}}\sin(\phi+\phi_S)\bigg)h^\nu_1 H^{\perp\nu}_1, \nonumber\\
\ee
which is contributed from azimuthal angle $\phi^h_q$ involved in fragmentation process. 
$\phi^h_q$ is the azimuthal angle of the produced hadron with respect to the 
fragmenting quark helicity frame and  
defined at $\mathcal{O}(\bfp^2/Q^2)$ 
as\cite{Anselmino:2011ch}
\be 
\cos\phi^h_q=\frac{P_{h\perp}}{k_\perp}\cos(\phi_h-\phi) - z \frac{p_\perp}{k_\perp}\\
\sin\phi^h_q=\frac{P_{h\perp}}{k_\perp}\sin(\phi_h-\phi).
\ee 
The pre-factors in the denominator and numerator of Eq.(\ref{Coll_Asy}) are the planar elementary hard cross-sections
\be
\frac{d\hat{\sigma}^{\ell q^\uparrow \to \ell q^\uparrow}}{dy}+\frac{d\hat{\sigma}^{\ell q^\uparrow \to \ell q^\downarrow}}{dy}
&=&\frac{d\hat{\sigma}}{dy}=\frac{2\pi \alpha^2}{s x y^2}[1+(1-y)^2],\\
\frac{d\hat{\sigma}^{\ell q^\uparrow \to \ell q^\uparrow}}{dy}-\frac{d\hat{\sigma}^{\ell q^\uparrow \to \ell q^\downarrow}}{dy}
&=&\frac{d(\Delta\hat{\sigma})}{dy}=\frac{4\pi \alpha^2}{s x y^2}[(1-y)].
\ee 
The Collins asymmetry defined in Eq.(\ref{Coll_Asy}) is a function of the variables $x, z, \bfPhp $ and $y$. 

Similarly, $\sin(3\phi_h-\phi_S)$ weighted azimuthal SSA $A^{\sin(3\phi_h-\phi_S)}_{UT}$ is defined as
\be 
A^{\sin(3\phi_h-\phi_S)}_{UT}(x,z,\bfPhp,y)
&=& \frac{4\pi^2\alpha^2\frac{(1-y)}{s x y^2} F^{\sin(3\phi_h-\phi_S)}_{UT}(x,z,\bfPhp,y) }
{2\pi^2\alpha^2\frac{1+(1-y)^2}{s x y^2}F_{UU}(x,z,\bfPhp,y)}.
\ee


The single spin asymmetry associated with longitudinally polarized proton is defined as
\be 
A_{UL}&=&\frac{d\sigma^{\ell P^\rightarrow \to \ell' h X}-d\sigma^{\ell P^\leftarrow \to \ell' h X}}{d\sigma^{\ell P^\rightarrow \to \ell' h X}+d\sigma^{\ell P^\leftarrow \to \ell' h X}} \nonumber\\
&=&\frac{d\sigma^{\ell P(S_L) \to \ell' h X}-d\sigma^{\ell P(-S_L)\to \ell' h X}}{d\sigma^{\ell P(S_L) \to \ell' h X}+d\sigma^{\ell P(-S_L) \to \ell' h X}}. 
\ee 
Where $\rightarrow, \leftarrow$ represent the longitudinal spin of proton along the momentum.
From Eq.(\ref{Cross}), the numerator of $A_{UL}$ can be written in terms of two structure functions $F^{\sin(2\phi_h)}_{UL} $ 
and $ F^{\sin(\phi_h)}_{UL} $   as 
\be 
\frac{d\sigma^{\ell P^\rightarrow \to \ell' h X}-d\sigma^{\ell P^\leftarrow \to \ell' h X}}{dx_B dy dz d^2\bfPhp d\phi_S}
&=& \frac{2\alpha^2}{s x y^2}2\bigg[(1-y)\sin(2\phi_h)F^{\sin(2\phi_h)}_{UL}+\sqrt{1-y}(2-y)\sin(\phi_h)F^{\sin(\phi_h)}_{UL}\bigg].\nonumber\\
\label{N_UT}
\ee
Where both the structure functions $F^{\sin(2\phi_h)}_{UL} $ and $ F^{\sin(\phi_h)}_{UL} $ have contribution from $h^{\perp\nu}_{1L}$ TMD 
and Collins FFs.
The associated asymmetries are given as
\be 
A^{\sin(2\phi_h)}_{UL}(x,z,\bfPhp,y)
&=& \frac{4\pi^2\alpha^2\frac{(1-y)}{s x y^2} F^{\sin(2\phi_h)}_{UL}(x,z,\bfPhp,y)  }{2\pi^2\alpha^2\frac{1+(1-y)^2}{s x y^2}F_{UU}(x,z,\bfPhp,y)},\\
A^{\sin(\phi_h)}_{UL}(x,z,\bfPhp,y)
&=& \frac{4\pi^2\alpha^2\frac{\sqrt{1-y}(2-y)}{s x y^2} F^{\sin\phi_h}_{UL}(x,z,\bfPhp,y)  }
{2\pi^2\alpha^2\frac{1+(1-y)^2}{s x y^2}F_{UU}(x,z,\bfPhp,y)}.
\ee

Using the TMDs from Eqs.(\ref{TMD_f1}-\ref{TMD_h1Tp}) and FFs from Eqs.(\ref{FF_D1},\ref{FF_H1}) into the Eqs.(\ref{SF_FUU}-\ref{SF_3hms}), the structure functions read in this model as 
\be 
F_{UU}&=&\sum_\nu e^2_\nu N^\nu_{f_1}\frac{\ln(1/x)}{\pi\kappa^2}\bigg[T^\nu_1(x)+
 \frac{\langle m^2_\perp \rangle}{M^2} T^\nu_2(x)\bigg]D^{h/\nu}_1(z) \avpsq_x \frac{e^{-\bfPhp^2/\avPhsq}}{\avPhsq}, \label{FUU_LFQDM} \\
F^{\sin(\phi_h+\phi_S)}_{UT}&=& \frac{P_{h\perp}\sqrt{2 e}}{M_h} \sum_\nu e^2_\nu  \hat{h}^{\nu}_1(x)\mathcal{N}^C_\nu(z) D^{h/\nu}_1(z) \frac{\avksq ^2_C \avpsq_x}{\avksq\avPhsq_C}\frac{e^{-\bfPhp^2/\avPhsq _C}}{\avPhsq_C}, \label{FCo_LFQDM} \\
F^{\sin(3\phi_h-\phi_S)}_{UT}&=& \frac{z^2 P^3_{h\perp}\sqrt{2 e}}{2 M_h M^2} \sum_\nu e^2_\nu  \hat{h}^{\perp\nu}_{1T}(x)\mathcal{N}^C_\nu(z) D^{h/\nu}_1(z) \frac{\avksq ^2_C \avpsq^3_x}{\avksq\avPhsq^3_C}\frac{e^{-\bfPhp^2/\avPhsq _C}}{\avPhsq_C}, \label{F3hms_LFQDM}\\
F^{\sin(2\phi_h)}_{UL}&=& \frac{z P^2_{h\perp}\sqrt{2 e}}{M_h M} \sum_\nu e^2_\nu  \hat{h}^{\perp\nu}_{1L}(x)\mathcal{N}^C_\nu(z) D^{h/\nu}_1(z) \frac{\avksq ^2_C \avpsq^2_x}{\avksq\avPhsq^2_C}\frac{e^{-\bfPhp^2/\avPhsq _C}}{\avPhsq_C}, \label{F2h_LFQDM}\\
F^{\sin(\phi_h)}_{UL}&=& -\frac{1}{Q}\frac{2 P_{h\perp}\sqrt{2 e}}{ M_h M} \sum_\nu e^2_\nu  \hat{h}^{\perp\nu}_{1L}(x)\mathcal{N}^C_\nu(z) D^{h/\nu}_1(z) \frac{\avksq ^2_C \avpsq^2_x}{\avksq\avPhsq^3_C}\nonumber\\
&&\hspace{1.5cm}\times\big[\avksq_C \avPhsq_C+ z \avpsq_x(P^2_{h\perp}-\avPhsq_C)\big]\frac{e^{-\bfPhp^2/\avPhsq _C}}{\avPhsq_C}. \label{Fh_LFQDM}
\ee

\begin{figure}
\includegraphics[width=6.2cm,clip]{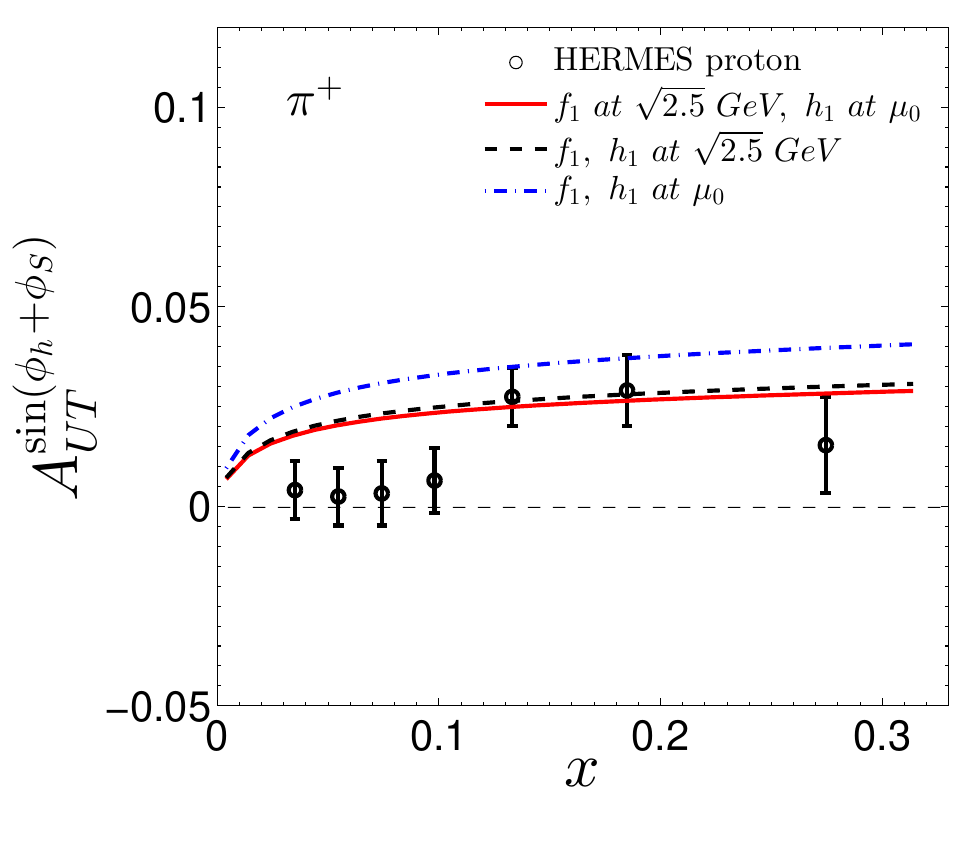} \includegraphics[width=6.2cm,clip]{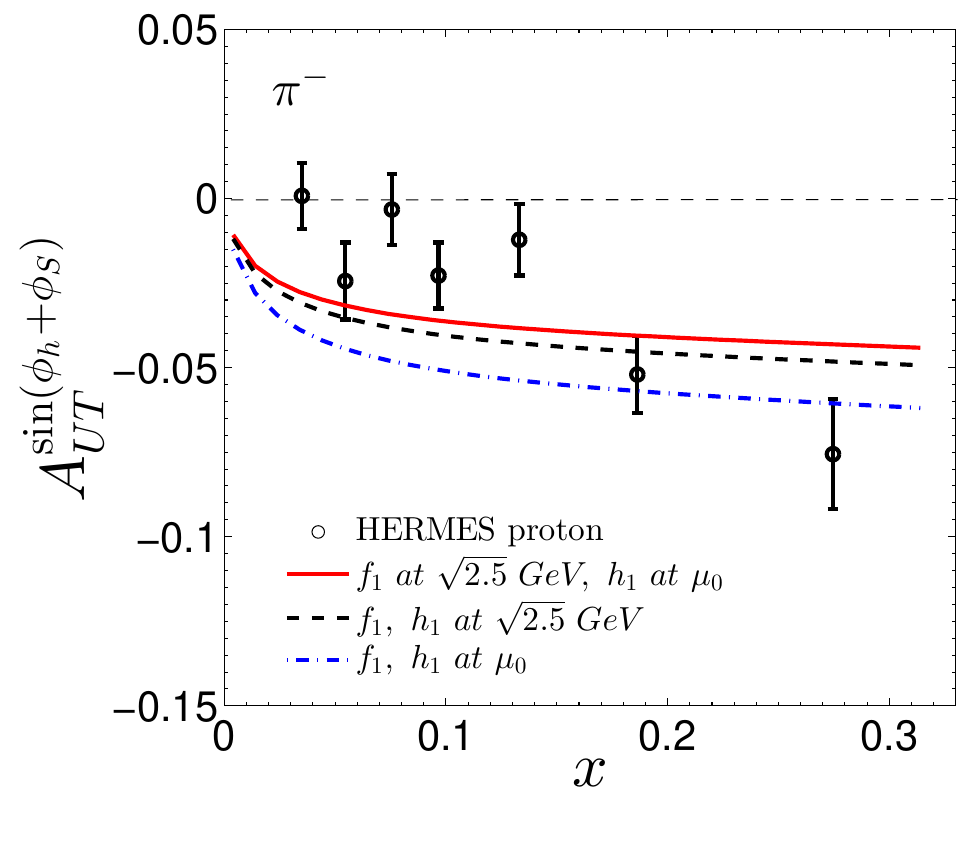} 
\caption{ \label{fig_stra} Variation of Collins asymmetry in the three strategies. Red continuous line: when $f_1$ is at $\mu^2=2.5~GeV^2$ and $h_1$ is at initial scale $\mu_0$. Black dashed line: when both $f_1$ and $h_1$ are at $\mu^2=2.5~GeV^2$. Blue dot-dashed line: when both $f_1$ and $h_1$ are at initial scale $\mu_0$. In all the cases, the TMDs are evolved in QCD evolution approach\cite{Aybat:2011zv,Anselmino:2012aa}, see Sec.\ref{evol}.}
\end{figure}

Thus, explicit expression of the single spin asymmetries, in LFQDM, are as the 
following:\\
(i) Collins asymmetry 
\be 
A^{\sin(\phi_h+\phi_S)}_{UT}&=&  \frac{\frac{2(1-y)}{s x y^2} \frac{P_{h\perp}\sqrt{2 e}}{M_h} \sum_\nu e^2_\nu  \hat{h}^{\nu}_1(x)\mathcal{N}^C_\nu(z) D^{h/\nu}_1(z)\frac{\avksq ^2_C \avpsq_x}{\avksq\avPhsq_C}\frac{e^{-\bfPhp^2/\avPhsq _C}}{\avPhsq_C}}{\frac{1+(1-y)^2}{s x y^2}\sum_\nu e^2_\nu N^\nu_{f_1}\frac{\ln(1/x)}{\pi\kappa^2}\bigg[T^\nu_1(x)+
 \frac{\langle m^2_\perp \rangle}{M^2} T^\nu_2(x)\bigg]D^{h/\nu}_1(z)\avpsq_x \frac{e^{-\bfPhp^2/\avPhsq}}{\avPhsq}}, \label{Coll_LFQDM}
\ee
(ii) the SSA $A^{\sin(3\phi_h-\phi_S)}_{UT}$
\be 
A^{\sin(3\phi_h-\phi_S)}_{UT}&=&  \frac{\frac{2(1-y)}{s x y^2}\frac{z^2 P^3_{h\perp}\sqrt{2 e}}{2 M_h M^2} \sum_\nu e^2_\nu  \hat{h}^{\perp\nu}_{1T}(x)\mathcal{N}^C_\nu(z) D^{h/\nu}_1(z) \frac{\avksq ^2_C \avpsq^3_x}{\avksq\avPhsq^3_C}\frac{e^{-\bfPhp^2/\avPhsq _C}}{\avPhsq_C}}{\frac{1+(1-y)^2}{s x y^2}\sum_\nu e^2_\nu N^\nu_{f_1}\frac{\ln(1/x)}{\pi\kappa^2}\bigg[T^\nu_1(x)+
 \frac{\langle m^2_\perp \rangle}{M^2} T^\nu_2(x)\bigg]D^{h/\nu}_1(z)\avpsq_x \frac{e^{-\bfPhp^2/\avPhsq}}{\avPhsq}}, \label{AUL_3h_LFQDM}
\ee
(iii) the SSA $A^{\sin(2\phi_h)}_{UL}$
\be 
A^{\sin(2\phi_h)}_{UL}&=&  \frac{\frac{2(1-y)}{s x y^2}\frac{z P^2_{h\perp}\sqrt{2 e}}{M_h M} \sum_\nu e^2_\nu  \hat{h}^{\perp\nu}_{1L}(x)\mathcal{N}^C_\nu(z) D^{h/\nu}_1(z) \frac{\avksq ^2_C \avpsq^2_x}{\avksq\avPhsq^2_C}\frac{e^{-\bfPhp^2/\avPhsq _C}}{\avPhsq_C}}{\frac{1+(1-y)^2}{s x y^2}\sum_\nu e^2_\nu N^\nu_{f_1}\frac{\ln(1/x)}{\pi\kappa^2}\bigg[T^\nu_1(x)+
 \frac{\langle m^2_\perp \rangle}{M^2} T^\nu_2(x)\bigg]D^{h/\nu}_1(z)\avpsq_x \frac{e^{-\bfPhp^2/\avPhsq}}{\avPhsq}}, \label{AUL_2h_LFQDM}
\ee
and (iv) the SSA $A^{\sin(\phi_h)}_{UL}$
\be 
A^{\sin(\phi_h)}_{UL}&=&  \frac{\frac{2(2-y)\sqrt{1-y}}{s x y^2} (-\frac{1}{Q})\frac{ P_{h\perp}2\sqrt{2 e}}{ M_h M} \sum_\nu e^2_\nu  \hat{h}^{\perp\nu}_{1L}(x)\mathcal{N}^C_\nu(z) D^{h/\nu}_1(z) \frac{\avksq ^2_C \avpsq^2_x}{\avksq\avPhsq^3_C} \langle \hat{m}^2_\perp \rangle \frac{e^{-\bfPhp^2/\avPhsq _C}}{\avPhsq_C}}{\frac{1+(1-y)^2}{s x y^2}\sum_\nu e^2_\nu N^\nu_{f_1}\frac{\ln(1/x)}{\pi\kappa^2}\bigg[T^\nu_1(x)+
 \frac{\langle m^2_\perp \rangle}{M^2} T^\nu_2(x)\bigg]D^{h/\nu}_1(z)\avpsq_x \frac{e^{-\bfPhp^2/\avPhsq}}{\avPhsq}}. \label{AUL_h_LFQDM}
\ee
Where,  
\be 
&\hat{h}^{\nu}_1(x)&= \bigg(C^2_SN^{\nu 2}_S-C^2_A\frac{1}{3}N^{\nu 2}_0\bigg) \frac{\ln(1/x)}{\pi\kappa^2}T^\nu_1(x)\\
&\hat{h}^{\nu\perp}_{1L}(x)&=-\bigg(C^2_SN^{\nu 2}_S+C^2_A\big(\frac{1}{3}N^{\nu 2}_0 - \frac{2}{3}N^{\nu 2}_1\big)\bigg)\frac{2\ln(1/x)}{\pi\kappa^2} T^\nu_3(x),\\
&\hat{h}^{\nu\perp}_{1T}(x)&= - \bigg(C^2_SN^{\nu 2}_S-C^2_A\frac{1}{3}N^{\nu 2}_0\bigg) \frac{2\ln(1/x)}{\pi\kappa^2}T^\nu_2(x),\\
&N^\nu_{f_1}&=C^2_SN^{\nu 2}_S+C^2_A\big(\frac{1}{3}N^{\nu 2}_0 + \frac{2}{3}N^{\nu 2}_1\big)\\
&\avpsq_x &= 1/R(x)= \frac{\kappa^2 (1-x)^2}{\delta \log(1/x)}\\
&\avksq_C &= \frac{M_h^2 \avksq}{M_h^2 + \avksq}\\
&\avPhsq_C & = \avksq_C + z^2 \avpsq_x\\
&\langle m^2_\perp \rangle &= \bigg[\avksq \avPhsq + z^2 P_{h\perp}^2 \avpsq_x\bigg]\frac{\avpsq_x}{\avPhsq^2}.\\
&\langle \hat{m}^2_\perp \rangle &=\bigg[\avksq_C \avPhsq_C+ z \avpsq_x(P^2_{h\perp}-\avPhsq_C )\bigg] \label{m_hat}.
\ee
The pre-factor $C_A$ represents $C_V$ and $C_{VV}$ for $u$ and $d$ quarks respectively. 
\begin{figure}
\includegraphics[scale=0.7]{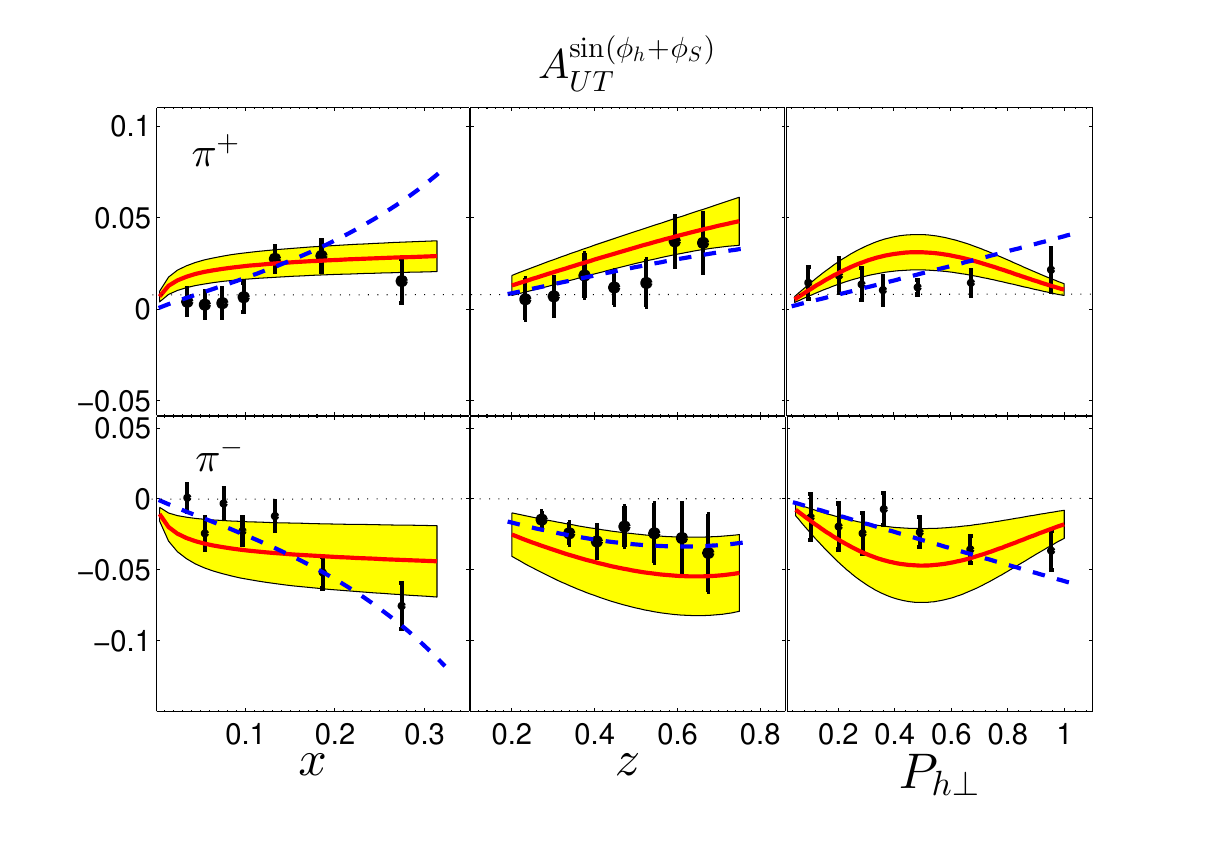} 
\caption{\label{fig_Col_H} Model prediction to Collins asymmetry  in SIDIS processes are presented and compared with experimental data 
by HERMES Collaboration\cite{Airapetian:2010ds}. Upper row and lower row are corresponding to $\pi^+$ and $\pi^-$ channels. First, second and 
third column represent the variation of asymmetry with respect to $x,z$ and $P_{h\perp}$. Red continuous lines(yellow error regions) 
represent the model result when $f^\nu_1$ is evolved in QCD evolution\cite{Aybat:2011zv,Anselmino:2012aa} at scale $\mu^2=2.5~GeV^2$. The blue dashed 
lines represent the model result when the TMDs are evolved in parameter evolution approach\cite{Maji:2016yqo}. In both  cases, $h_1$ remains at 
the initial scale and FFs are taken from the parametrization\cite{Anselmino:2013vqa,Kretzer:2001pz} at $\mu^2=2.5~GeV^2$.}
\end{figure}


\begin{figure}
\includegraphics[scale=0.7]{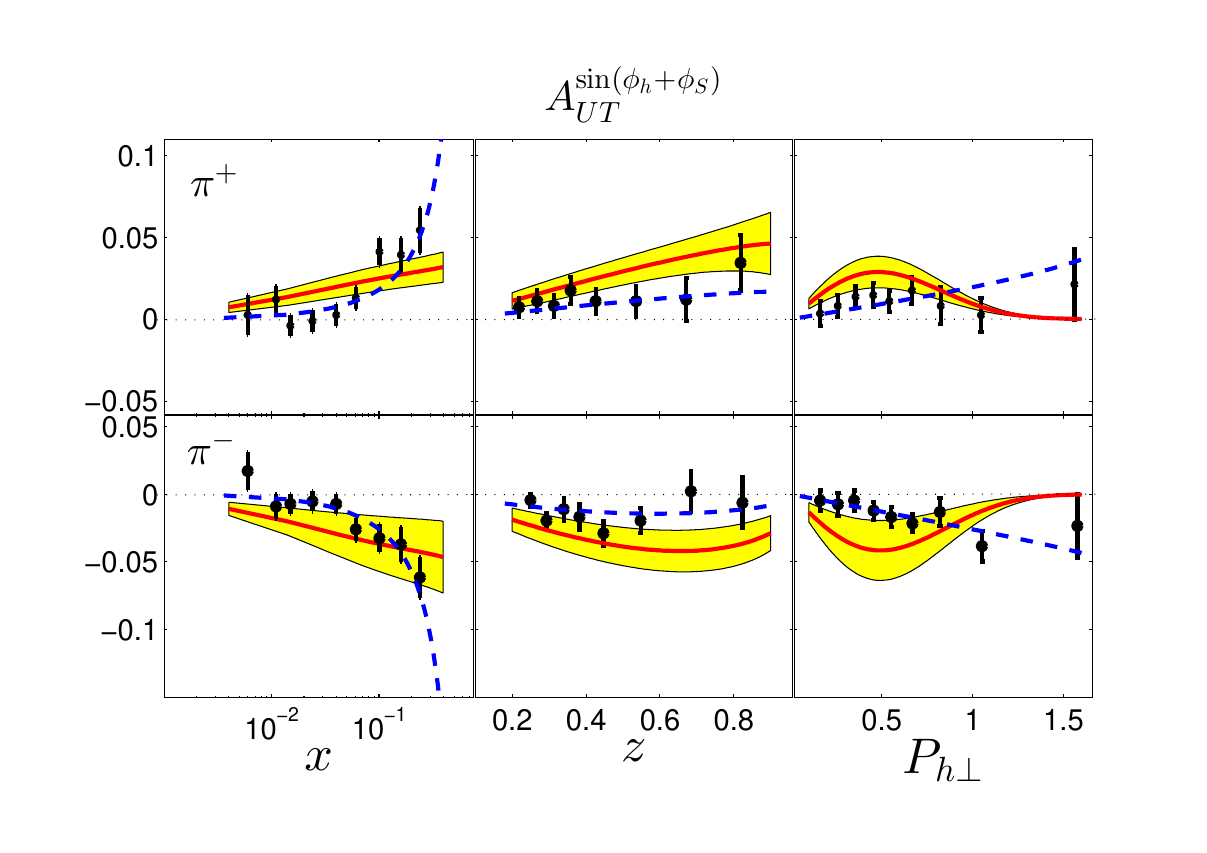}
\caption{\label{fig_Col_C} Model prediction to Collins asymmetry  measured by COMPASS Collaboration\cite{Martin:2013eja} for $\pi^+$ and $\pi^-$ channels. Red continuous lines(yellow error regions) and blue dashed lines represent the same as of Fig.\ref{fig_Col_H}.}
\end{figure} 


\subsection{Predictions for COMPASS and HERMES}
All the above asymmetries are functions of $x,z,\bfPhp,y$ and scale $\mu$ whereas the experimental measurements of asymmetries provide 
the variation of the integrated asymmetry with one variable at a time. Therefore one has to integrate the denominator and numerator separately 
over all the other variables except that one variable which is measured in that data. Also to compare with the experimental data it is needed 
to keep the $x,y,z$ dependence canceling factors in the numerator and denominator of asymmetries unchanged. 

An amount of integrated asymmetry can be estimated by integrating over all the variables $x,z,P_{h\perp}$ and $y$ in the corresponding 
kinametical limits i.e.,
\be 
\!\!\tilde{A}^{\mathcal{W}(\phi_h,\phi_S)}_{S_\ell S_P}&=&2\frac{\int dx dz dP_{h\perp} dy \bigg(\int d\phi_h d\phi_S [d\sigma^{\ell(S_\ell)
P(S_P) \to \ell' h X}-d\sigma^{\ell(S_\ell) P(-S_P) \to \ell' h X}]\mathcal{W}(\phi_h,\phi_S)\bigg)}{\int dx dz dP_{h\perp} 
dy \bigg(\int d\phi_h d\phi_S [d\sigma^{\ell(S_\ell) P(S_P) \to \ell' h X}+d\sigma^{\ell(S_\ell) P(-S_P) \to \ell' h X}]\bigg)}. \label{Asy_Int}
\ee
The kinematical limit for the variables in the HERMES experiment are:\\
\be 
0.023 \leq x \leq 0.4,~~ 0.2 \leq z \leq 0.7  ~~ and ~~ 0.1 \leq y \leq 0.95 
\ee
  and in the COMPASS experiment are:
\be 
0.003 \leq x \leq 0.7,~~ 0.2 \leq z \leq 1.0 ~~ and ~~ 0.1 \leq y \leq 0.9
\ee
The model result for the amount of the integrated asymmetries are listed in the table.\ref{tab_IA} for both the $\pi^+$ and $\pi^-$ channels. The experimental 
data are available only for different values of kinematical variables so 
direct comparison is not possible, the signs of different asymmetries evaluated 
in the model 
are consistent with the data.
The amplitude of asymmetries are calculated following the same strategy i.e., $f^\nu_1$ is evolved in QCD evolution at $\mu^2=2.5~GeV^2$ 
and the polarized TMDs involved in the numerator remains at the initial scale. 
\begin{table}[h]
\centering  
\begin{tabular}{|c c c c c |}
 \hline
 Channel~~&~~$\tilde{A}^{\sin(\phi_h+\phi_S)}_{UT}|_{HERMES}$~~&~~$\tilde{A}^{\sin(\phi_h+\phi_S)}_{UT}|_{COMPASS}$~~ & $\tilde{A}^{\sin(3\phi_h-\phi_S)}_{UT}$~~ & ~~ $\tilde{A}^{\sin\phi_h }_{UL}$   \\ \hline
$\pi^+$ ~~&~~ 0.0236 ~~&~~ 0.0374 ~~&~~ -0.0011 ~~&~~ 0.0336 \\
$\pi^-$ ~~&~~ -0.0364 ~~&~~ -0.0534 ~~&~~ 0.0015 ~~&~~ -0.0518  \\ \hline
\end{tabular} 
\caption{Amount of the integrated SSAs in this model, from Eq.(\ref{Asy_Int}), 
for both the $\pi^+$ and $\pi^-$ channels. The amplitudes 
$\tilde{A}^{\sin(3\phi_h-\phi_S)}_{UT}$ and $\tilde{A}^{\sin \phi_h}_{UT}$ are 
calculated in HERMES kinematics.  }  
\label{tab_IA}  
\end{table}

Since the COMPASS and HERMES experiments measure the asymmetries at higher scale($\mu>\mu_0$), scale evolution of the asymmetries are
needed to give a model prediction to the experimental data. Scale evolution of the associated TMDs and FFs provide the scale evolution 
of the asymmetries. Therefore, effectively we need to evolve the TMDs and FFs. But well defined scale evolutions for all the leading twist TMDs, 
specially for chiral-odd TMDs, are not known yet. A QCD evolution of unpolarized TMDs, $f^\nu_1$, and FFs, $D^{h/\nu}_1$, is 
discussed in \cite{Aybat:2011zv, Aybat:2011ge}(see Sec.\ref{evol}).A reduced form of QCD evolution is proposed in \cite{Anselmino:2012aa} and adopted to the 
evolution of spin-dependent TMDs e.g., Sivers functions. Similarly one can adopt 
this QCD evolution  for all the polarized 
TMDs and predict  the asymmetries. To understand qualitatively, we compare our 
result for Collins asymmetry in the three different 
schemes(shown in Fig.\ref{fig_stra}): (i) $f^\nu_1$ is at $\mu^2=2.5~GeV^2$ and $h^\nu_1$ is at initial scale, 
(ii) both $f^\nu_1$ and $h^\nu_1$  are at $\mu^2=2.5~GeV^2$ and (iii) $f^\nu_1$ and $h^\nu_1$ are at the initial scale $\mu_0^2$. Interestingly, the 
scheme-(i) gives better result among these three schemes. The evolution 
contribution from $h^\nu_1$ is very small for Collins asymmetry(scheme-(ii)). 
Note that, in the case of other asymmetries e.g., $A^{\sin(3\phi_h-\phi_S)}$, 
the scheme-(ii) has a large deviation from the data. 
Therefore we evolve the unpolarized TMDs, $f^\nu_1$, which is known and contributes to the denominator of the asymmetries and all spin-dependent 
TMDs whose evolutions are not well known and are involved in numerators of all the asymmetries are taken at initial scale.  
Not only this strategy gives better agreement with data but limits the 
uncertainty to the numerators of the asymmetries only. A similar strategy is used in\cite{Boffi:2009sh}.

The average bin energy range for the HERMES is $1.3 ~GeV^2 < Q^2 < 6.2~GeV^2 $ and the average $Q^2$ values of the HERMES 
experiment is around $2.4 ~GeV^2$. We use the LO parametrization for $D^{h/\nu}_1$ and $H^{\perp \nu}_1$ at the 
scale $2.5 ~GeV^2$ \cite{Kretzer:2001pz,Anselmino:2012aa}. So, we evolve $f^\nu_1$ to the same scale ($\mu^2=2.5 ~GeV^2$) to give a model prediction 
for Collins asymmetry as well as for other azimuthal asymmetries. 

We perform evolution of $f^\nu_1$ in two different approaches: one is the QCD evolution approach as discussed in Sec.\ref{evol} and 
another one is the TMDs evolution by parameter evolution approach of LFQDM proposed in \cite{Maji:2016yqo}. In the parameter evolution approach,
the parameters in the LFQDM are allowed to evolve to generate the DGLAP evolution of the unpolarized PDFs. The same evolution of the parameters
are used to estimate the TMD evolution.
So,  the information of DGLAP evolution are  encoded into the parameters and the TMDs  are expected to follow more like DGLAP evolution in 
this approach.
   
\begin{figure}
\includegraphics[scale=0.7]{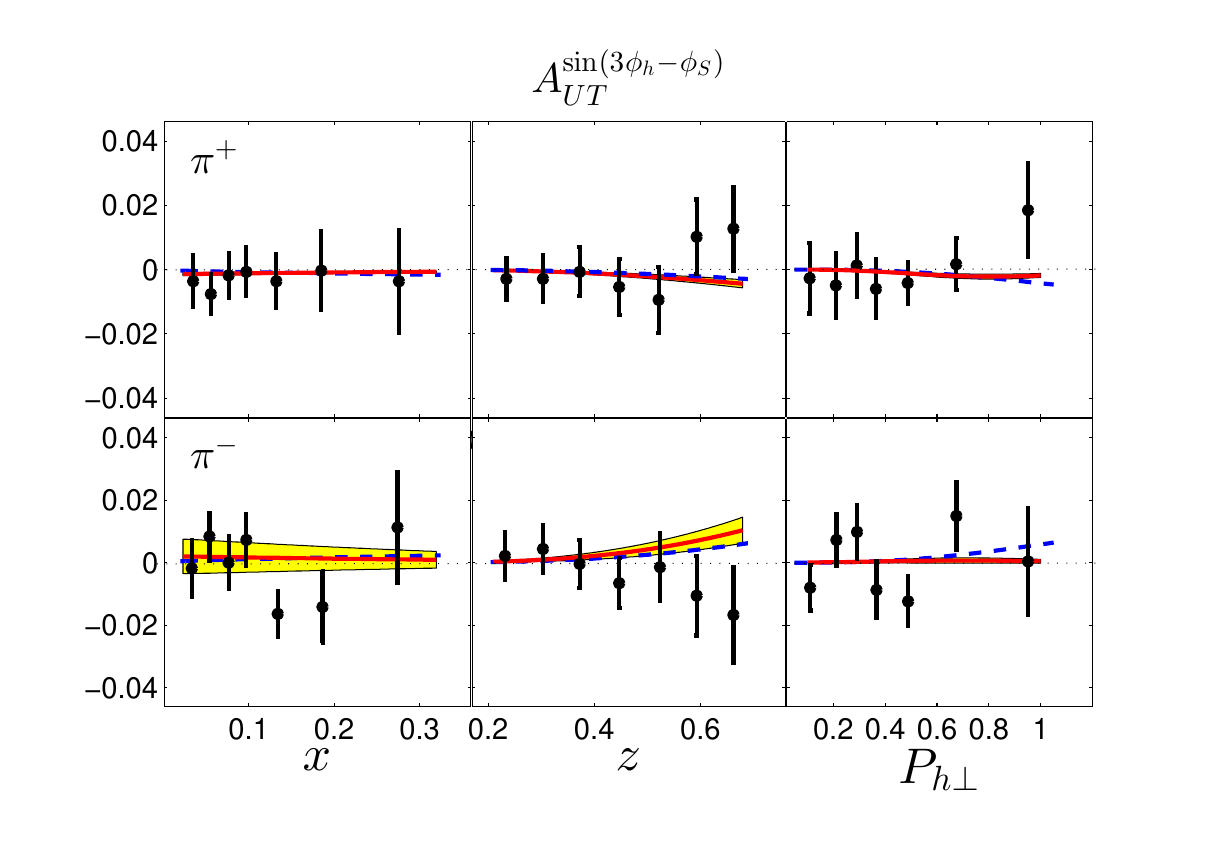}
\caption{ \label{fig_A3hms} Model prediction to the SSA 
$A^{\sin(3\phi_h-\phi_S)}_{UT}$ for $\pi^+$(upper row) and $\pi^-$(lower row) 
channels. 
The first, second and third column represent the $x, z$ and $P_{h\perp}$ 
variations respectively. Red continuous lines(yellow error region) 
and blue dashed lines represent the model prediction to this SSA when $f^\nu_1$ is evolved in QCD evolution and in parameter evolution 
respectively at the scale $\mu^2=2.5~GeV^2$ . In both the cashes, $h^{\perp\nu}_{1T}$ are at initial scale and FFs are taken as 
phenomenological input at the scale $\mu^2=2.5~GeV^2$. Data are measured by HERMES\cite{Schnell:2010zza}. } 
\end{figure}


\begin{figure}
\includegraphics[scale=0.7]{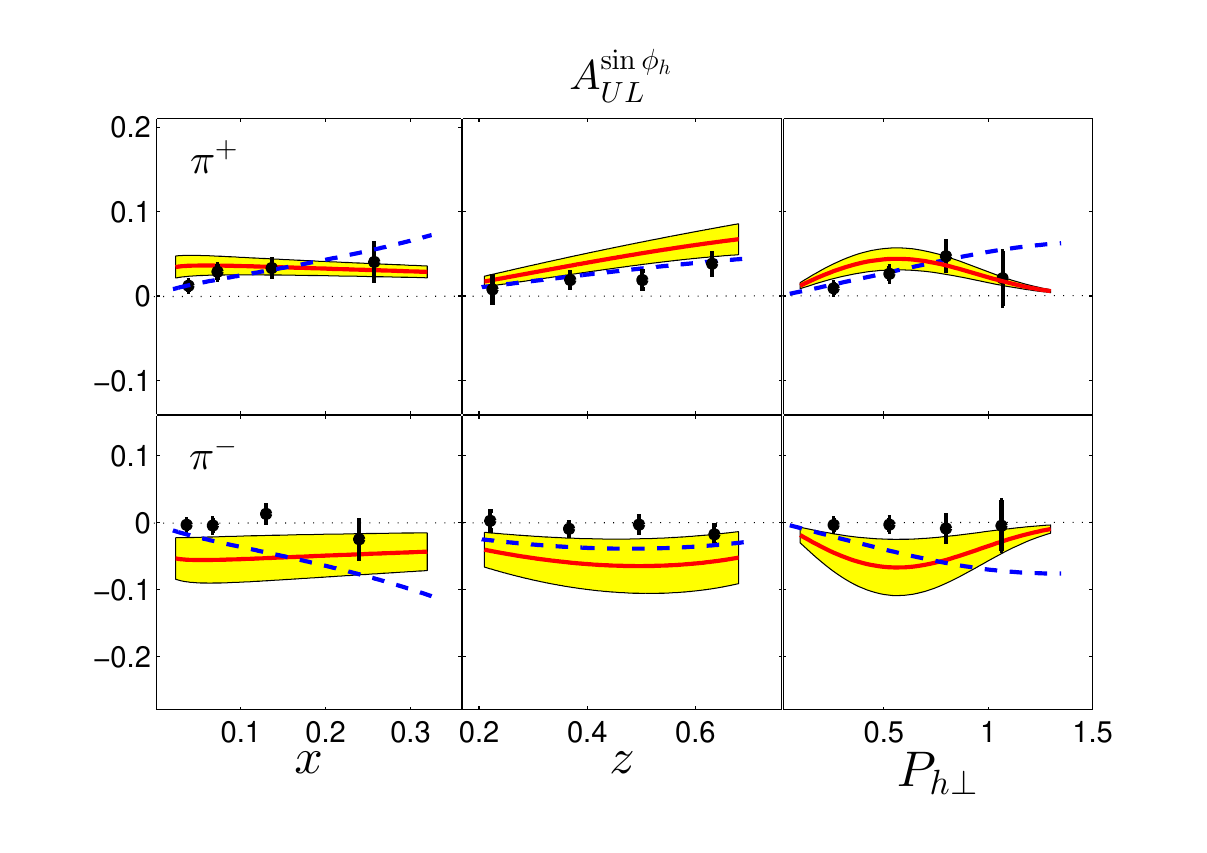} 
\caption{ \label{fig_sinh} Model prediction to $ A^{\sin \phi_h}_{UL}$ for 
$\pi^+$(upper row) and $\pi^-$(lower row) channels. The first, second and third 
column represent the $x, z$ and $P_{h\perp}$ variations respectively. 
The colors and symbols have the
same interpretations as in Fig.\ref{fig_Col_H}. Data are measured by HERMES 
collaboration\cite{Airapetian:2002mf}.}
\end{figure}



Our model predictions to the Collins asymmetry are shown in Fig.\ref{fig_Col_H} and compared with the HERMES data for the 
kinematics $0.023 \leq x \leq 0.4,~~ 0.2 \leq z \leq 0.7 $ and $0.1 \leq y \leq 0.95$. The upper row is for $\pi^+$ and the 
lower row is for $\pi^-$ production channels. The first, second and third columns indicate the $x, z$ and $\bfPhp$ variations 
of Collins asymmetry respectively. The red continuous lines represent the model prediction of Collins asymmetry where the $f^\nu_1$ 
is evolved in QCD evolution given in\cite{Aybat:2011zv,Anselmino:2012aa}, see Sec.\ref{evol}. The corresponding error is represented in yellow color.  
The error corridors are coming from the uncertainties in the parameters of TMDs(initial scale error) and FFs. Error coming from the LFQDM 
is small,  large contributions come from the uncertainties in the parameters of 
FFs\cite{Anselmino:2013vqa}.
The model predicts qualitative behavior of the asymmetries 
and   agree with the data within error bar and we expect that when QCD 
evolution of all the TMDs and FFs are correctly incorporated, the agreement 
with the data will improve. 
The blue dashed lines represent the model
prediction when the TMD $f_1$  is evolved by parameter evolution approach\cite{Maji:2016yqo}. Error corridor for blue dashed line is not shown to avoid 
clumsiness in the plot. 
Since a well defined QCD evolution for transversity is not available, we evolve 
the unpolarized TMD only and 
restrict the uncertainty to the numerator of Collins asymmetry. Using the same strategy in parameter evolution approach 
we observe a fantastic agreement to the experimental data(denoted by blue dashed 
line). The model results of Collins asymmetry 
for $\pi^+$ channel and $\pi^-$ channel are positive and negative respectively as found in experimental measurements. In this model, 
the amount of the Collins asymmetries(in the HERMES kinematics) are 0.0236 and 
 -0.0364 for $\pi^+$ and $\pi^-$ channels respectively 
(see Table \ref{tab_IA}). 

In Fig.{\ref{fig_Col_C}, the model result for Collins asymmetry is compared with the COMPASS data corresponding to the 
kinematics: $0.003 \leq x \leq 0.7,~~ 0.2 \leq z \leq 1.0 $ and $0.1 \leq y \leq 
0.9$.  All the colors and indicators represent 
the same as used in Fig.\ref{fig_Col_H}. We observe that our model prediction to 
the Collins asymmetry is quite reasonable.
As for HERMES, the agreement of the model predictions for 
variation with $P_{h\perp}$ is not so good. The parameter evolution 
approach(blue line) again shows excellent agreement with the COMPASS data.
In this model, 
the amount of integrated asymmetries(in the COMPASS kinematics) are 0.0374 and 
-0.0534 in $\pi^+$ and $\pi^-$ channels respectively(see Table 
\ref{tab_IA}).

  


Model prediction to the single spin asymmetry $A^{\sin(3\phi_h-\phi_S)}_{UT}$ is shown in Fig.\ref{fig_A3hms} and compared with 
HERMES data\cite{Schnell:2010zza}. (The color and signs of plots represent the same 
as of Fig.\ref{fig_Col_H}.) This asymmetry involves 
pretzelocity distribution and characterizes the $\bfp$ dependence of the transverse quark polarization in a transversely polarized proton. 
The pretzelocity TMD is linked to the non spherical shape of the proton and
quark orbital angular momentum. 
Compared to $A_{UT}^{\sin(\phi_h+\phi_S)}$, this 
asymmetry is suppressed by a factor of 
 $P_{h\perp}^2/M^2$ and hence expected to be very small for small transverse 
momentum 
of the outgoing hadron $\mid P_{h\perp}\mid <M$, where $M$ is the proton mass  
 (see Eq.(\ref{AUL_3h_LFQDM})).
 Experimental results show that the 
asymmetries as functions
of $x, z$ or $P_{h\perp}$  are near equal to zero as shown 
in Fig.\ref{fig_A3hms}. Our model results also predict
almost negligible asymmetries for both the channels. 
As a result, the amount of integrated asymmetries(Eq.(\ref{Asy_Int})) are also 
very small and are found to be  -0.0011 and 0.0015 for the $\pi^+$ and $\pi^-$ 
channels respectively.


The model prediction to SSA $A^{\sin\phi_h}_{UL}$ is shown in Fig.\ref{fig_sinh} and compared with the HERMES data\cite{Airapetian:2002mf} for 
$\pi^+$ and $\pi^-$ production channels. The colors and symbols represent the same as in Fig.\ref{fig_Col_H}. This asymmetry has 
contribution from $h^{\perp \nu}_{1L}(x,\bfp^2)$ TMD, see 
Eq.(\ref{AUL_h_LFQDM}). 
 Compared to $A_{UT}^{\sin(\phi_h+\phi_S)}$ (Eq.(\ref{Coll_LFQDM})), the 
asymmetry $A_{UL}^{\sin{\phi_h}}$ is  suppressed by a factor 
of $P_{h\perp}/M$ and expected to be smaller than Collins asymmetry for small 
transverse momenta of the outgoing hadron which is consistent with the 
available data. Model result for the amplitude of the asymmetries are 0.0336 
and -0.0518 in $\pi^+$ and $\pi^-$ channels respectively.  

Note that the parameter evolution is a model to reproduce the DGLAP 
evolution of the PDFs, but it is found to work well to reproduce the SSAs too. 
The TMDs are known  not to follow the DGLAP evolution, and the same parameter 
evolution  is not expected to reproduce their evolutions.  But in the SSAs, 
which involve ratios of different TMDs and fragmentation functions, it  
seems to work fine which might be due to partial cancellations of the evolution 
effects. Proper QCD evolutions of all the TMDs and FFs are required for more 
accurate predictions of the asymmetries at the experimental scales.


\subsection{Prediction for EIC}
\begin{figure}
\centering
\includegraphics[width=15.0cm,clip]{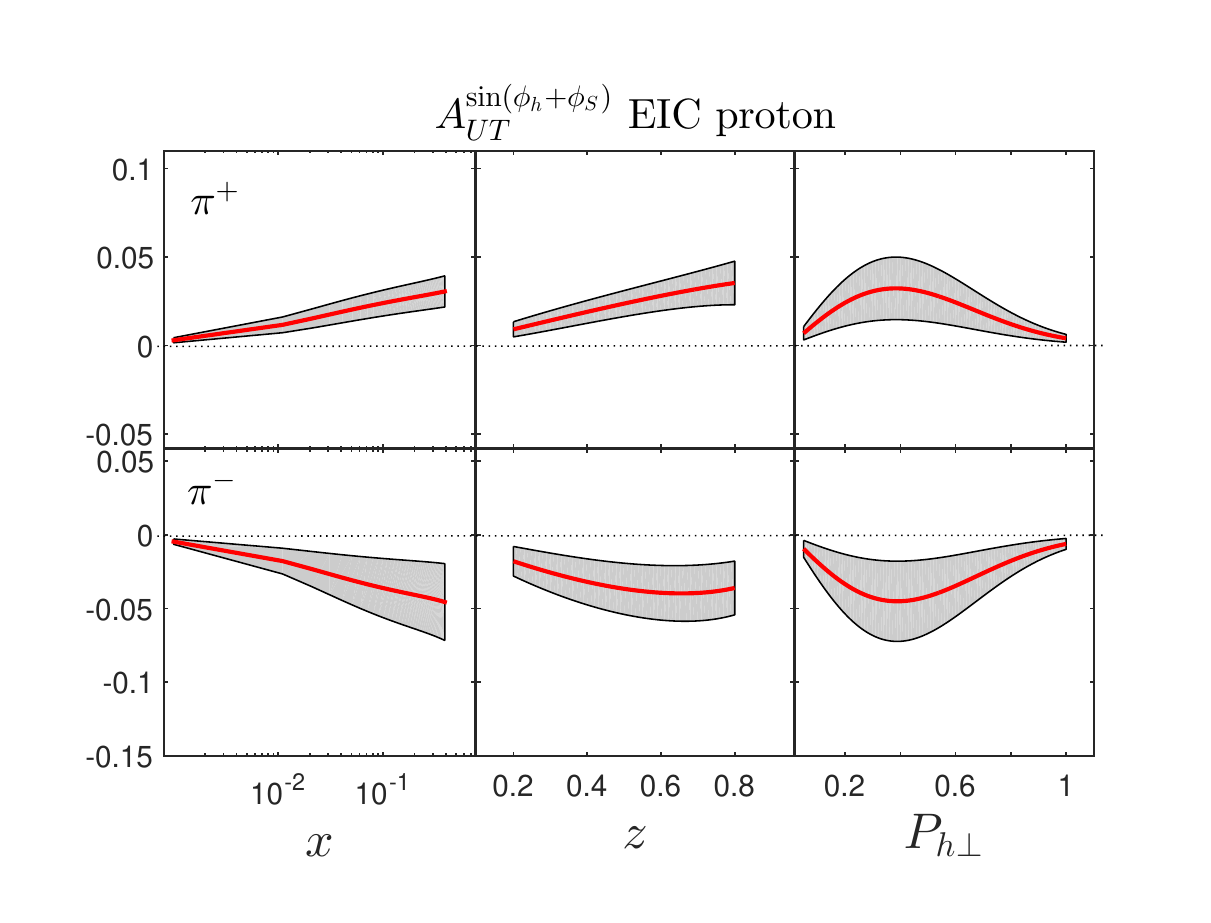}
\caption{ Prediction for the Collins asymmetry $A_{UT}^{\sin(\Phi_h+\Phi_S)}$ for EIC kinematics.\label{eic_fig}}
\end{figure}
The upcoming Electron Ion Collider(EIC)\cite{Accardi:2012qut} is designed to use 
 several existing facilities to
 probe both DIS and SIDIS over a wide range of kinematics and beam polarization.
It is expected to provide much deeper insight into the hadron structure. Here we 
present the model predictions
for the  Collins asymmetry  
for the EIC kinematics.  We present our predictions for the EIC 
kinematics\cite{Wang:2016tix}:
\be
0.001 < x < 0.4, && 0.2 < z < 0.8,\\ 0.05 < {P_h}_\perp < 1, && 0.01 < y < 0.95,
\ee
at the center of mass energy 
$\sqrt{s}=45$ GeV. The predictions for the collins asymmetry $A_{UT}^{\sin(\Phi_h+\Phi_S)}$ at $\mu^2=100~ GeV^2$ 
are shown in Fig.\ref{eic_fig}. Note that the future EIC will explore much smaller values of $x$ as can be seen from the plots. The upper panel in 
Fig.\ref{eic_fig} represents the results for $\pi^+$ channel while the
lower panel is for $\pi^-$ channel and the asymmetries are predicted to be 
sizable in both channels.

\section{Double Spin Asymmetries in LFQDM}\label{sec_DSAs}
The double-spin asymmetry is observed when both the lepton beam and the target proton are polarized and only proton polarization flips.
The DSAs associated with the  longitudinally polarized lepton beam is defined as
\be 
A_{LS_P}&=&\frac{d\sigma^{\ell^\rightarrow P(S_P) \to \ell' h X}-d\sigma^{\ell^\rightarrow P(S_P) \to \ell' h X}}{d\sigma^{\ell^\rightarrow P(S_P) \to \ell' h X}+d\sigma^{\ell^\rightarrow P(S_P) \to \ell' h X}} \nonumber\\
\ee
Where, the target proton can considered as longitudinally polarized($S_L \equiv 
\rightarrow $) or transversely polarized  ($S_T \equiv \uparrow$). For 
longitudinally polarized proton, from Eq.(\ref{Cross}), the numerator can be 
written in terms of the structure functions as
\be 
\frac{d\sigma^{\ell^\rightarrow P^\rightarrow \to \ell' h X}-d\sigma^{\ell^\rightarrow P^\leftarrow \to \ell' h X}}{dx_B dy dz d^2\bfPhp d\phi_S}&=& 
\frac{2\alpha^2}{s x y^2}2\bigg[\frac{1-(1-y)^2}{2}F_{LL}+y\sqrt{1-y}\cos\phi_h F^{\cos\phi_h}_{LL}\nonumber\\
&+&(1-y)\sin2\phi_h F^{\sin2\phi_h}_{UL}+ (2-y)\sqrt{1-y}\sin\phi_h F^{\sin\phi_h}_{UL}\bigg].
\label{N_LL}
\ee

Where the first two structure functions contribute to the double spin asymmetries. The DSAs for longitudinally polarized proton and lepton beam are defined in terms of structure functions as
\be
A_{LL}(x,z,\bfPhp,y)&=& \frac{2\pi^2\alpha^2\frac{1-(1-y)^2}{s x y^2} F_{LL}(x,z,\bfPhp,y)  }{2\pi^2\alpha^2\frac{1+(1-y)^2}{s x y^2}F_{UU}(x,z,\bfPhp,y)},\\
A^{\cos(\phi_h)}_{LL}(x,z,\bfPhp,y)&=&  \frac{4\pi^2\alpha^2\frac{y\sqrt{1-y}}{s x y^2} F^{\cos(\phi_h)}_{LL}(x,z,\bfPhp,y)  }{2\pi^2\alpha^2\frac{1+(1-y)^2}{s x y^2}F_{UU}(x,z,\bfPhp,y)}.
\ee

The double spin asymmetry with longitudinally polarized lepton and transversely polarized proton is defined as
\be 
A_{LT}&=&\frac{d\sigma^{\ell^\rightarrow P^\uparrow \to \ell' h X}-d\sigma^{\ell^\rightarrow P^\downarrow \to \ell' h X}}{d\sigma^{\ell^\rightarrow P^\uparrow \to \ell' h X}+d\sigma^{\ell^\rightarrow P^\downarrow \to \ell' h X}} \nonumber\\
\ee
From Eq.(\ref{Cross}),the numerator is written as
\be 
\frac{d\sigma^{\ell^\rightarrow P^\uparrow \to \ell' h X}-d\sigma^{\ell^\rightarrow P^\downarrow \to \ell' h X}}{dx_B dy dz d^2\bfPhp d\phi_S} &=& \frac{2\alpha^2}{s x y^2}2\bigg[\frac{1-(1-y)^2}{2}\cos(\phi_h-\phi_S)F^{\cos(\phi_h-\phi_S)}_{LT} \nonumber\\
&+& y\sqrt{1-y}\bigg(\cos\phi_S F^{\cos\phi_S}_{LT} + \cos(2\phi_h-\phi_S)F^{\cos(2\phi_h-\phi_S)}_{LT}\bigg)\bigg],
\ee
and the weighted DSAs for transversely polarized proton are given by
\be 
A^{\cos(\phi_h-\phi_S)}_{LT}(x,z,\bfPhp,y) &=& \frac{2\pi^2\alpha^2\frac{1-(1-y)^2}{s x y^2} F^{\cos(\phi_h-\phi_S)}_{LT}(x,z,\bfPhp,y)  }{2\pi^2\alpha^2\frac{1+(1-y)^2}{s x y^2}F_{UU}(x,z,\bfPhp,y)},\\
A^{\cos\phi_S}_{LT}(x,z,\bfPhp,y) &=& \frac{4\pi^2\alpha^2\frac{y\sqrt{1-y}}{s x y^2} F^{\cos\phi_h}_{LT}(x,z,\bfPhp,y)}{2\pi^2\alpha^2\frac{1+(1-y)^2}{s x y^2}F_{UU}(x,z,\bfPhp,y)},\\
A^{\cos(2\phi_h-\phi_S)}_{LT}(x,z,\bfPhp,y) &=& \frac{4\pi^2\alpha^2\frac{y\sqrt{1-y}}{s x y^2} F^{\cos(2\phi_h-\phi_S)}_{LT}(x,z,\bfPhp,y)}{2\pi^2\alpha^2\frac{1+(1-y)^2}{s x y^2}F_{UU}(x,z,\bfPhp,y)}
\ee



\begin{figure}
\includegraphics[scale=0.7]{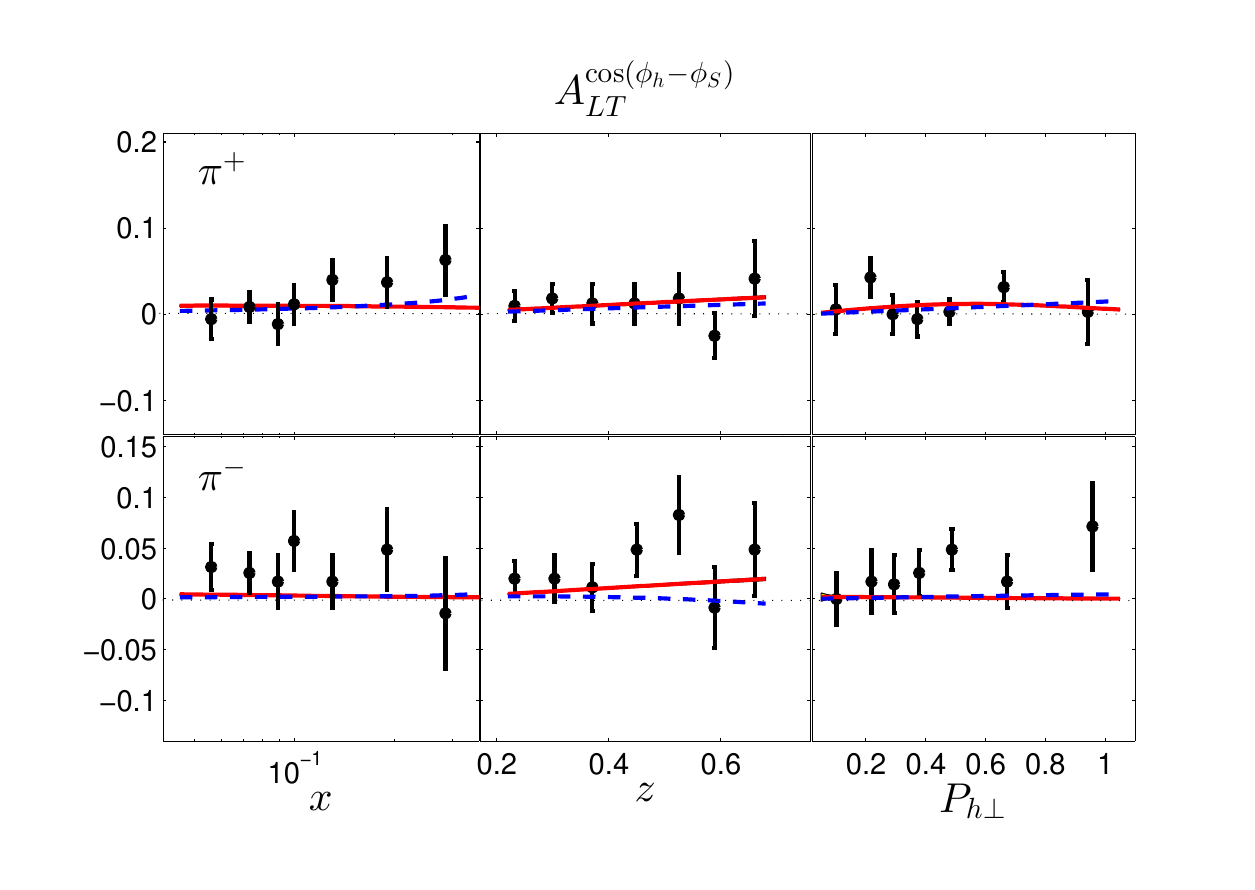}
\caption{\label{fig_DSA_hms} Model prediction to double spin asymmetry $A^{\cos(\phi_h-\phi_S)}_{LT}$ for proton are shown and compared with HERMES data\cite{Pappalardo:2014sga,Pappalardo:2012zz} for $\pi^+$ and $\pi^-$ channels. Red lines(yellow error region) and blue dashed lines represent the model prediction to this DSA when $f^\nu_1$ is evolved in QCD evolution and in parameter evolution respectively. $g^\nu_{1T}$ is at initial scale and FFs are taken as phenomenological input at the scale $\mu^2=2.5~GeV^2 $.} 
\end{figure}


The  structure functions  in this model read as 
\be 
F_{LL}&=&\sum_\nu e^2_\nu N^\nu_{g_1}\frac{\ln(1/x)}{\pi\kappa^2}\bigg[T^\nu_1(x)-
 \frac{\langle m^2_\perp \rangle}{M^2} T^\nu_2(x)\bigg]D^{h/\nu}_1(z) \avpsq_x \frac{e^{-\bfPhp^2/\avPhsq}}{\avPhsq},\\
 F^{\cos\phi_h}_{LL}&=&(-\frac{2}{Q})z \Pht \sum_\nu e^2_\nu N^\nu_{g_1}\frac{\ln(1/x)}{\pi\kappa^2}\bigg[T^\nu_1(x)-
 \frac{\langle m^2_\perp \rangle}{M^2} T^\nu_2(x)\bigg]D^{h/\nu}_1(z) \frac{(\avpsq _x)^2}{\avPhsq} \frac{e^{-\bfPhp^2/\avPhsq}}{\avPhsq},\\ 
 F^{\cos(\phi_h-\phi_S)}_{LT}&=& \frac{z \Pht}{M} \sum_\nu e^2_\nu  \hat{g}^{\nu}_{1T}(x) D^{h/\nu}_1(z) \frac{\big(\avpsq_x\big)^2}{\avPhsq}\frac{e^{-\bfPhp^2/\avPhsq}}{\avPhsq},\\
 F^{\cos\phi_S}_{LT}&=& \bigg(-\frac{1}{Q}\bigg)\frac{1}{M}\sum_\nu e^2_\nu  \hat{g}^{\nu}_{1T}(x) D^{h/\nu}_1(z) \frac{(\avpsq_x)^2}{(\avPhsq)^2}\bigg[\avksq \avPhsq + z^2 \Pht^2 \avpsq \bigg]\frac{e^{-\bfPhp^2/\avPhsq}}{\avPhsq},\\
  F^{\cos(2\phi_h-\phi_S)}_{LT}&=& \bigg(-\frac{1}{Q}\bigg)\frac{z^2 \Pht^2}{M}\sum_\nu e^2_\nu  \hat{g}^{\nu}_{1T}(x) D^{h/\nu}_1(z) \frac{(\avpsq_x)^3}{(\avPhsq)^2}\frac{e^{-\bfPhp^2/\avPhsq}}{\avPhsq}. 
\ee

\begin{figure}
\includegraphics[scale=0.85]{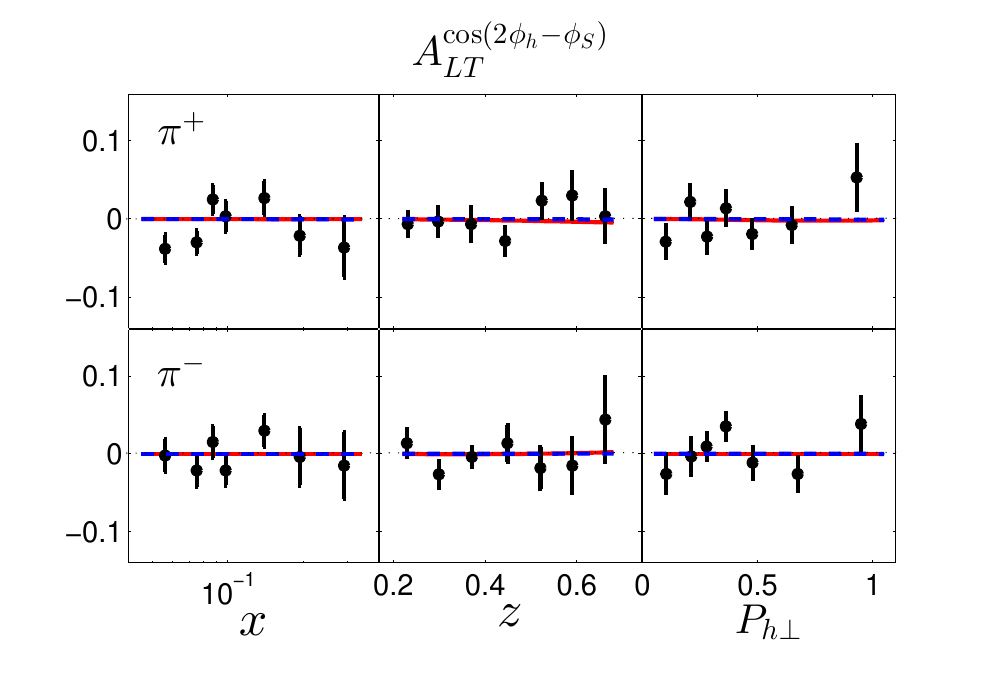} 
\caption{ \label{fig_DSA_cos2hms} Model prediction to $ A^{\cos(2\phi_h-\phi_S)}_{LT}$ for $\pi^+$(upper row) and $\pi^-$(lower row) channels are shown and compared with the preliminary HERMES data \cite{Pappalardo:2012zz}. The first, second and third column represent the $x, z$ and $P_{h\perp}$ variations respectively. The red continuous lines(yellow error region) indicates the same as in Fig.\ref{fig_DSA_hms}.}
\end{figure}

Thus, in this model the explicit form of the double spin asymmetries are given by
\be 
A_{LL}&=&\frac{\frac{1-(1-y)^2}{s x y^2} \sum_\nu e^2_\nu N^\nu_{g_1}\frac{\ln(1/x)}{\pi\kappa^2}\bigg[T^\nu_1(x)-
 \frac{\langle m^2_\perp \rangle}{M^2} T^\nu_2(x)\bigg]D^{h/\nu}_1(z) \avpsq_x \frac{e^{-\bfPhp^2/\avPhsq}}{\avPhsq}}{\frac{1+(1-y)^2}{s x y^2}\sum_\nu e^2_\nu N^\nu_{f_1}\frac{\ln(1/x)}{\pi\kappa^2}\bigg[T^\nu_1(x)+
 \frac{\langle m^2_\perp \rangle}{M^2} T^\nu_2(x)\bigg]D^{h/\nu}_1(z) \avpsq_x \frac{e^{-\bfPhp^2/\avPhsq}}{\avPhsq}}\\
 A^{\cos\phi_h}_{LL}&=&\frac{\frac{2y\sqrt{1-y}}{s x y^2} (-\frac{2}{Q})z \Pht \sum_\nu e^2_\nu N^\nu_{g_1}\frac{\ln(1/x)}{\pi\kappa^2}\bigg[T^\nu_1(x)-
 \frac{\langle m^2_\perp \rangle}{M^2} T^\nu_2(x)\bigg]D^{h/\nu}_1(z) \frac{(\avpsq _x)^2}{\avPhsq} \frac{e^{-\bfPhp^2/\avPhsq}}{\avPhsq}}{\frac{1+(1-y)^2}{s x y^2}\sum_\nu e^2_\nu N^\nu_{f_1}\frac{\ln(1/x)}{\pi\kappa^2}\bigg[T^\nu_1(x)+
 \frac{\langle m^2_\perp \rangle}{M^2} T^\nu_2(x)\bigg]D^{h/\nu}_1(z) \avpsq_x \frac{e^{-\bfPhp^2/\avPhsq}}{\avPhsq}}\\
  A^{\cos(\phi_h-\phi_S)}_{LT}&=&\frac{\frac{1-(1-y)^2}{s x y^2} \frac{z \Pht}{M} \sum_\nu e^2_\nu  \hat{g}^{\nu}_{1T}(x) D^{h/\nu}_1(z) \frac{(\avpsq_x)^2}{\avPhsq}\frac{e^{-\bfPhp^2/\avPhsq}}{\avPhsq}}{\frac{1+(1-y)^2}{s x y^2}\sum_\nu e^2_\nu N^\nu_{f_1}\frac{\ln(1/x)}{\pi\kappa^2}\bigg[T^\nu_1(x)+
 \frac{\langle m^2_\perp \rangle}{M^2} T^\nu_2(x)\bigg]D^{h/\nu}_1(z) \avpsq_x \frac{e^{-\bfPhp^2/\avPhsq}}{\avPhsq}}\\
  A^{\cos\phi_S}_{LT}&=&\frac{\frac{2y\sqrt{1-y}}{s x y^2} (-\frac{1}{Q})\frac{1}{M}\sum_\nu e^2_\nu  \hat{g}^{\nu}_{1T}(x) D^{h/\nu}_1(z) \frac{(\avpsq_x)^2}{(\avPhsq)^2}\bigg[\avksq \avPhsq + z^2 \Pht^2 \avpsq \bigg]\frac{e^{-\bfPhp^2/\avPhsq}}{\avPhsq}}{\frac{1+(1-y)^2}{s x y^2}\sum_\nu e^2_\nu N^\nu_{f_1}\frac{\ln(1/x)}{\pi\kappa^2}\bigg[T^\nu_1(x)+
 \frac{\langle m^2_\perp \rangle}{M^2} T^\nu_2(x)\bigg]D^{h/\nu}_1(z) \avpsq_x \frac{e^{-\bfPhp^2/\avPhsq}}{\avPhsq}}\\
 A^{\cos(2\phi_h-\phi_S)}_{LT}&=&\frac{\frac{2y\sqrt{1-y}}{s x y^2} (-\frac{1}{Q})\frac{z^2 P^2_{h\perp}}{M}\sum_\nu e^2_\nu  \hat{g}^{\nu}_{1T}(x) D^{h/\nu}_1(z) \frac{(\avpsq_x)^3}{(\avPhsq)^2}\frac{e^{-\bfPhp^2/\avPhsq}}{\avPhsq}}{\frac{1+(1-y)^2}{s x y^2}\sum_\nu e^2_\nu N^\nu_{f_1}\frac{\ln(1/x)}{\pi\kappa^2}\bigg[T^\nu_1(x)+ \frac{\langle m^2_\perp \rangle}{M^2} T^\nu_2(x)\bigg]D^{h/\nu}_1(z) \avpsq_x \frac{e^{-\bfPhp^2/\avPhsq}}{\avPhsq}}.
\ee
Where,
\be 
\hat{g}^{\nu}_{1T}(x)&=& \bigg(C^2_SN^{\nu 2}_S-C^2_A\frac{1}{3}N^{\nu 2}_0\bigg)\frac{2\ln(1/x)}{\pi\kappa^2} T^\nu_3(x),\\
N^\nu_{g_1}&=&\bigg(C^2_SN^{\nu 2}_S+C^2_A\big(\frac{1}{3}N^{\nu 2}_0-\frac{2}{3}N^{\nu 2}_1\big)\bigg).
\ee

Here all the DSAs are functions of $x, z, \bfPhp, y$ at a scale $\mu$. The DSAs $A_{LL}$ and $A^{\cos\phi_h}_{LL}$ have contribution from the 
helicity TMD, $g^\nu_{1L}$ . The other three DSAs $ A^{\cos(\phi_h-\phi_S)}_{LT}, A^{\cos \phi_S}_{LT}$ and $ A^{\cos(\phi_h-\phi_S)}_{LT}$ have 
contributions from the worm-gear TMD, $g^\nu_{1T}$.

The model prediction of $\cos(\phi_h-\phi_S)$ weighted double spin asymmetry $ A^{\cos(\phi_h-\phi_S)}_{LT}$ for longitudinally polarized lepton 
and transversely polarized proton are shown in Fig.\ref{fig_DSA_hms}. The error bar is very small in this case and  presented by yellow region. 
 Our results show reasonably good agreement with the HERMES data. This asymmetry is found to be slightly positive for both 
the $\pi^+$ and $\pi^-$ channels as observed by HERMES experiment\cite{Pappalardo:2012zz,Pappalardo:2014sga}. Positive asymmetry for $\pi^-$ channel is also found 
in Hall-A results on transversely polarized $3He$ target. In our model, amount 
of  the integrated asymmetries are very small, 0.0093 and 0.0032 for $\pi^+$ 
and $\pi^-$ 
channels respectively (see table.\ref{tab_IDSA}) and are consistent with the 
experimental data. 

The model result for DSA $ A^{\cos(2\phi_h-\phi_S)}_{LT}$ is shown in 
Fig.\ref{fig_DSA_cos2hms}. The colors and notations are the same as in
Fig.\ref{fig_Col_H}. The data are taken from HERMES measurement\cite{Pappalardo:2012zz}. In 
the HERMES measurement, this asymmetry is found to be nearly equal to zero for 
both the $\pi^+$ and $\pi^-$ channels. Our model also shows almost zero 
asymmetry for $x$ variation, whereas a slight positive asymmetry is observed 
for the case of $P_{h\perp}$ variation. 
Note that the model error is very small and  presented by 
yellow region. The amount of integrated asymmetries 
are given in Table.\ref{tab_IDSA}.

\begin{table}[h]
\centering  
\begin{tabular}{|c c c |}
 \hline
 Channel~~&~~$\tilde{A}^{\cos(\phi_h-\phi_S)}_{LT}$~~&~~$\tilde{A}^{\cos(2\phi_h-\phi_S)}_{LT}$   \\ \hline
$\pi^+$ ~~&~~ 0.0093 ~~&~~ -0.00033  \\
$\pi^-$ ~~&~~ 0.0032 ~~&~~ -0.00006   \\ \hline
\end{tabular} 
\caption{Amount of DSAs in this model, from Eq.(\ref{Asy_Int}), for both the $\pi^+$ and $\pi^-$ channels. The amplitudes
$\tilde{A}^{\cos(\phi_h-\phi_S)}_{LT}$ and $\tilde{A}^{\cos(2\phi_h-\phi_S)}_{LT}$ are calculated in HERMES kinematics. }  
\label{tab_IDSA}  
\end{table}


\begin{figure}[htbp]
\includegraphics[width=7.2cm,clip]{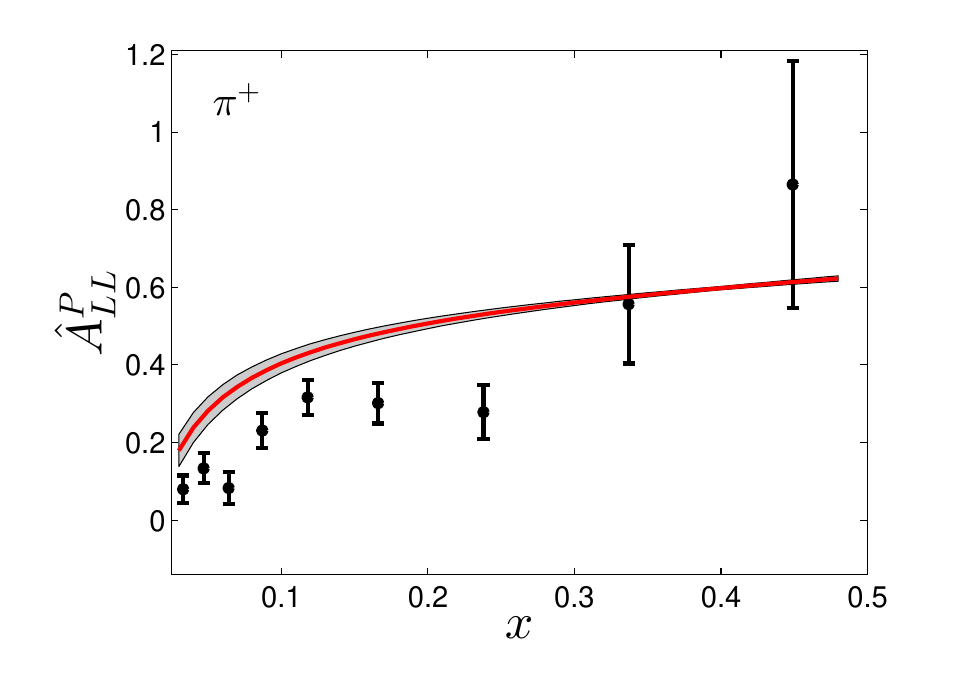}
\includegraphics[width=7.2cm,clip]{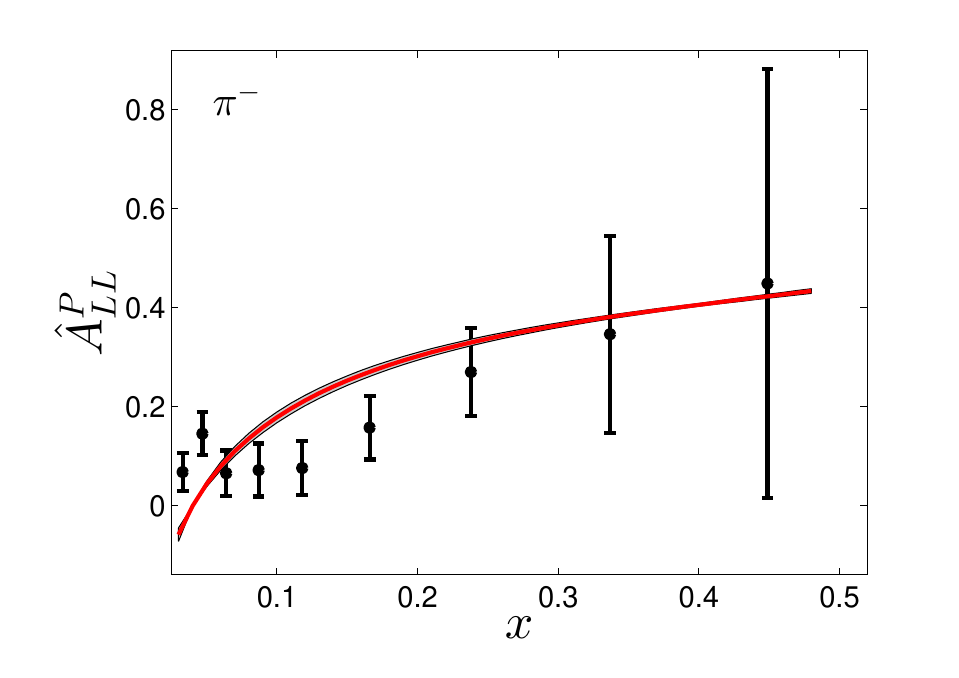}
\caption{\label{fig_ALL_xz} Model prediction to $A^p_{LL}$ for proton, at scale $\mu^2=2.5~GeV^2$, is shown by red continuous line. The left plot is for $\pi^+$ channel and the right plot is for $\pi^-$ channel at $ z = 0.46$. The data are taken from \cite{Airapetian:2004zf}.}
\end{figure}

In the SIDIS process, the integrated DSA(integrated over transverse 
momentum) $\hat{A}^P_{LL}(x,z,\mu)$ is measured by the HERMES collaboration and 
defined in terms of helicity PDFs as
\be
\hat{A}^P_{LL}(x,z,\mu)=\frac{\sum_\nu e^2_\nu g_1(x,\mu)  D^{h/\nu}_1(z,\mu)}{\sum_\nu e^2_\nu f_1(x,\mu) D^{h/\nu}_1(z,\mu)}\label{ALL_xz}
\ee
The model result for $x$ variation of $ \hat{A}^P_{LL}(x,z,\mu) $  are shown in 
Fig.\ref{fig_ALL_xz} and compared with the HERMES result \cite{Airapetian:2004zf} for  
$\pi^+$ and $\pi^-$ channels. We have taken the bin average values for $z=0.46$ 
in the HERMES experiment. All the distributions in Eq.(\ref{ALL_xz}) are taken 
at the scale $\mu^2=2.5~GeV^2$. Since the parameter evolution is consistent 
with the DGLAP evolution, the helicity PDF and unpolarized PDF are evolved in 
parameter evolution approach.


\begin{figure}[htbp]
\includegraphics[width=7.2cm,clip]{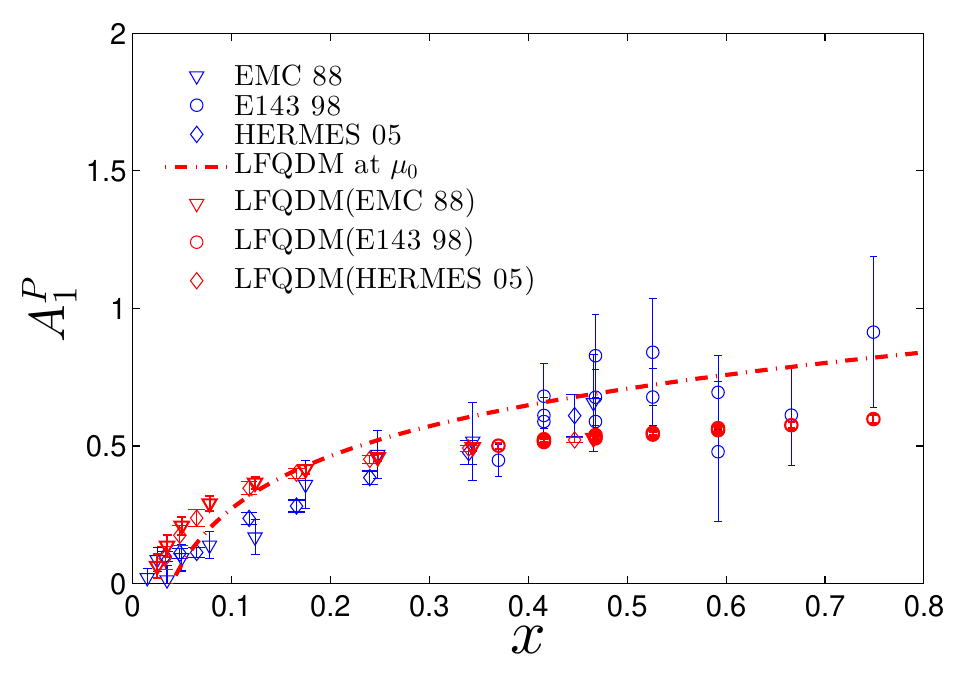}
\caption{\label{fig_A1} Model prediction to double spin asymmetry $A^P_1(x)$ for proton: red data are the model predictions corresponding to the measured scale $\mu$. The experimental data are taken from \cite{Ashman:1989ig, Abe:1998wq, Airapetian:2004zf} and denoted by the blue color with experimental error bar. The red dash doted line represents the asymmetry when all the distributions($f_1$ and $g_1$) are at initial scale $\mu_0$. }
\end{figure}

If no hadron is observed in the final state, the double spin asymmetry for proton is given by
\be
A^P_1=\frac{\sum_\nu e^2_\nu g_1(x)}{\sum_\nu e^2_\nu f_1(x)}
\ee
which have the contribution from PDFs only (no 
contribution from FFs). 
In this model, the variation of $A^P_1$ with $x$ is shown in Fig.\ref{fig_A1} and 
compared with the experimental data\cite{Ashman:1989ig, Abe:1998wq, Airapetian:2004zf}. The red dot-dashed 
line represents the asymmetry when both the PDFs $f_1(x)$ and $h_1(x)$ are at 
initial scale $\mu_0$. The red data points represent the model result 
corresponding to the set of $x$ and $\mu$ values measured experimentally at EMC, 
E134 and HERMES\cite{Ashman:1989ig, Abe:1998wq, Airapetian:2004zf}. Since $A_1$ symmetry involves the PDFs, 
we use the parameter evolution approach(which is consistent with DGLAP 
evolution) for the scale evolution. Since the evolutions of the PDFs are 
well known, as expected the model predictions are in good agreement with the 
data.

\section{Relations}
From the Fig.\ref{fig_R_Soff} we can write a model dependent inequality as
\be 
A^{\sin(\phi_h+\phi_s)}_{UT}(P_{h\perp}) \leq \frac{1}{2}|A_{UU}(P_{h\perp})+A_{LL}(P_{h\perp})|
\ee
The above inequality can be considered as a Soffer bound type relation for asymmetries, which provides an upper cut for Collins asymmetry in SIDIS process.
\begin{figure}[htbp]
\includegraphics[width=7.2cm,clip]{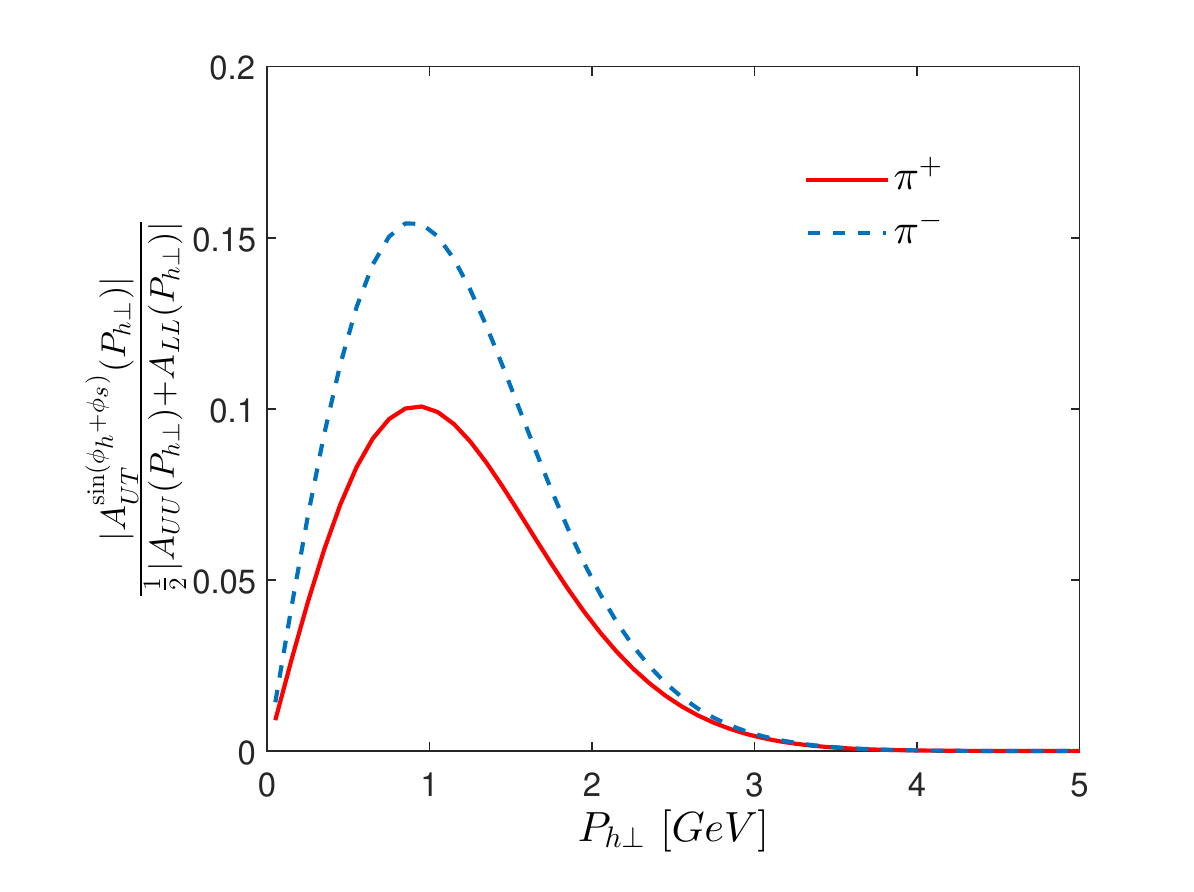}
\caption{\label{fig_R_Soff} Ratio of the asymmetries $\frac{A^{\sin(\phi_h+\phi_s)}_{UT}(P_{h\perp})}{\frac{1}{2}|A_{UU}(P_{h\perp})+A_{LL}(P_{h\perp})|}$ .}
\end{figure}

Similarly Fig.\ref{fig_R_Pretz} provides an upper bound for $A^{\sin(3\phi_h-\phi_s)}_{UT}(P_{h\perp})$ as
\be 
|\frac{\bfPhp^2}{2 M^2} A^{\sin(3\phi_h-\phi_s)}_{UT}(P_{h\perp})| \leq \frac{1}{2}|A_{UU}(P_{h\perp})-A_{LL}(P_{h\perp})|
\ee
\begin{figure}[htbp]
\includegraphics[width=7.2cm,clip]{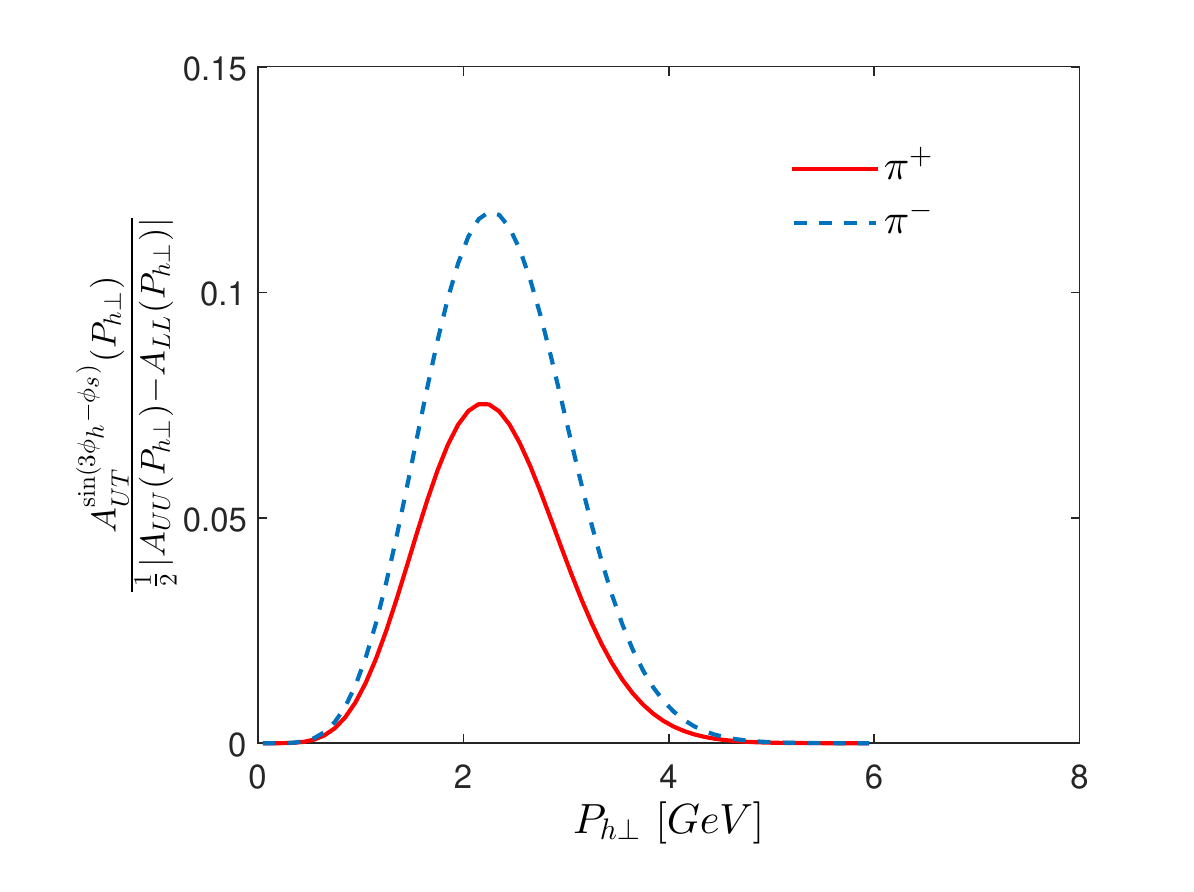}
\caption{\label{fig_R_Pretz} Ratio of the asymmetries $\frac{A^{\sin(3\phi_h-\phi_s)}_{UT}(P_{h\perp})}{\frac{1}{2}|A_{UU}(P_{h\perp})-A_{LL}(P_{h\perp})|}$ .}
\end{figure}

\begin{figure}[htbp]
\includegraphics[width=7.2cm,clip]{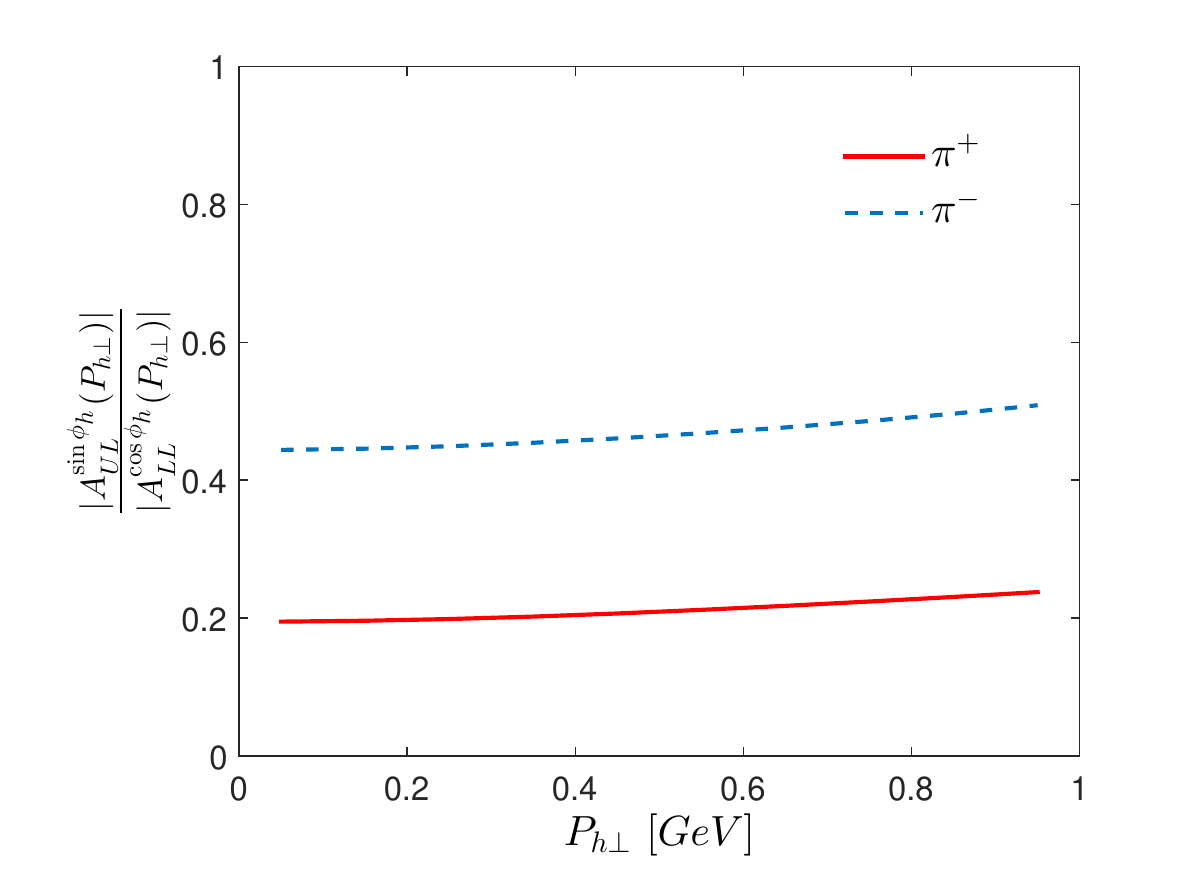}
\includegraphics[width=7.2cm,clip]{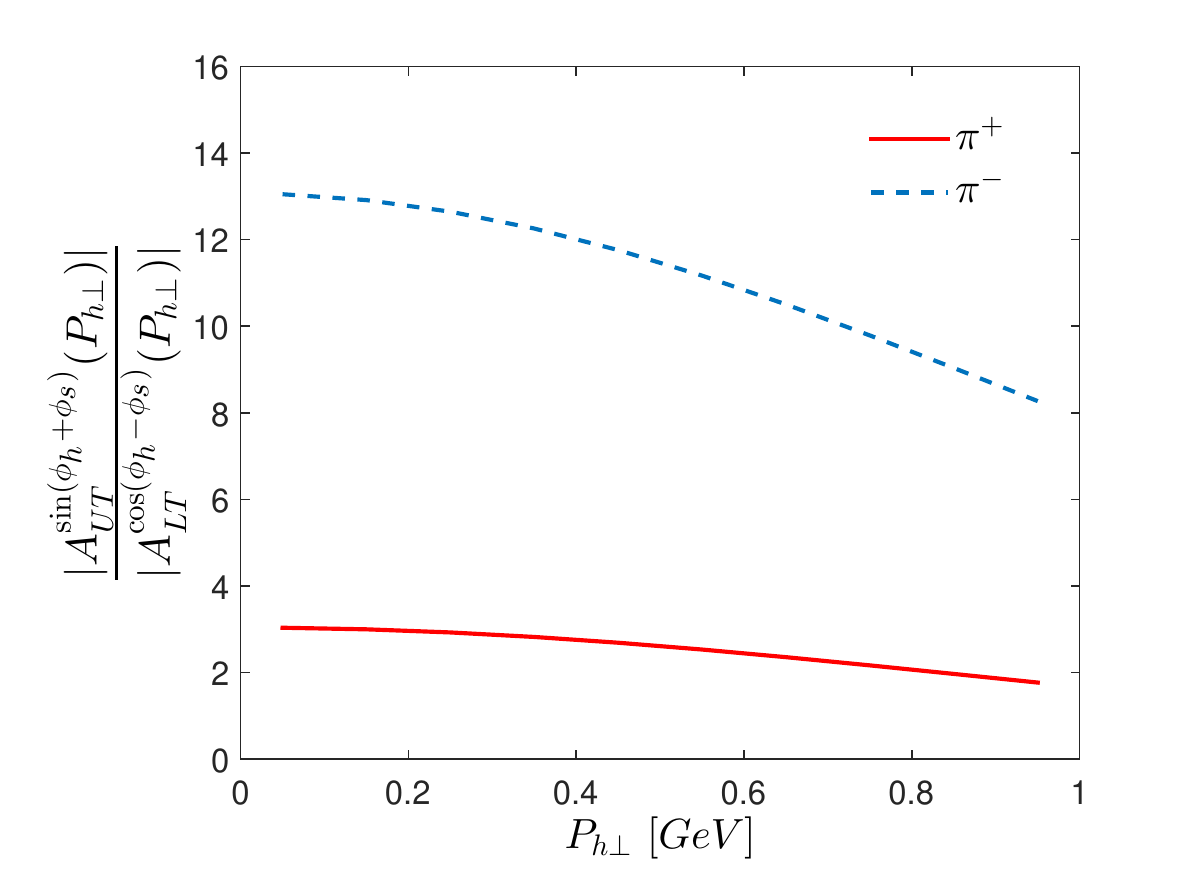}
\caption{\label{fig_R_asym} Ratio of the asymmetries.}
\end{figure}

In this model, relations among the SSAs and DSAs can be written as
\be
\frac{A^{\sin(2\phi_h)}_{UL}\big/(z P_{h\perp})}{A^{\sin(\phi_h)}_{UL} \avPhsq _C \big/\langle \hat{m}^2_\perp \rangle} &=& (-Q) \frac{1-y}{2(2-y)\sqrt{1-y}}\\
\frac{A_{LL}}{A^{\cos \phi_h}_{LL} \avPhsq \big/ (z P_{h\perp} \avpsq _x)} &=&(-Q)\frac{1-(1-y)^2}{4y\sqrt{1-y}}\\
\frac{A^{\cos(\phi_h-\phi_S)}_{LT}\big/(z P_{h\perp})}{A^{\cos \phi_S}_{LT} \avPhsq \big/\langle \hat{n}^2_\perp \rangle} &=&(-Q)\frac{1-(1-y)^2}{2 y \sqrt{1-y}}\\
\frac{A^{\cos(\phi_h-\phi_S)}_{LT}}{A^{\cos(2\phi_h-\phi_S)}_{LT} \avPhsq\big/(z P_{h\perp} \avpsq _x) } &=& (-Q) \frac{1-(1-y)^2}{2 y \sqrt{1-y}}
\ee
Where $\langle \hat{n}^2_\perp \rangle=[\avksq \avpsq + z^2 P^2_{h\perp} \avpsq]$ and $\langle \hat{m}^2_\perp \rangle$ is given in the Eq.(\ref{m_hat}). Right hand side of all these equations are functions of $y$ and $Q$ only and independent of $x,z$ and $P_{h \perp}$.

The magnitudes of the ratios of DSA over SSA e.g, $ | 
A_{UL}^{\sin(\phi_h)}(P_{h\perp})/A_{LL}^{\cos(\phi_h)}(P_{h\perp})| $ or $ 
|A_{UT}^{\sin(\phi_h+\phi_S)}(P_{h\perp})/A_{LL}^{\cos(\phi_h-\phi_S)}(P_{
h\perp} )|$  for $\pi^+$ channel are found to be smaller than the $\pi^-$ 
channel as shown in Fig.\ref{fig_R_asym}. 
One of the possible reasons for this result is that apart from many other 
factors,  
the SSA/DSA ratios involve the ratio of the fragmentation functions 
$H_1^\perp(k_\perp)/D_1(k_\perp)$ and  this ratio of the fragmentation 
functions   for $u$ quark is smaller than the same for 
 $d$ quark(Fig.\ref{fig_R_FF}). Since $u\to \pi^+$ and $d\to \pi^-$ are the 
favored fragmentations, it suggests that the SSA/DSA ratio for $\pi^+$ channel 
should be smaller than the $\pi^-$ channel.
Note that, the ratio of fragmentation function accounts for about a factor 
of $1.5$ whereas 
$|A_{UL}^{\sin(\phi_h)}(P_{h\perp})/A_{LL}^{\cos(\phi_h)}(P_{h\perp})| $ 
for $\pi^-$ is about twice that for $\pi^+$ channel and the ratio  $ 
|A_{UT}^{\sin(\phi_h+\phi_S)}(P_{h\perp})/A_{LL}^{\cos(\phi_h-\phi_S)}(P_{
h\perp} )|$ for $\pi^-$ is bigger than the $\pi^-$ channel by a factor of 
almost four. So, not only the fragmentation functions but  other  
quantities like difference in the values of TMDs for $u$ and $d$  also play 
important role.

\begin{figure}[htbp]
\includegraphics[width=7.2cm,clip]{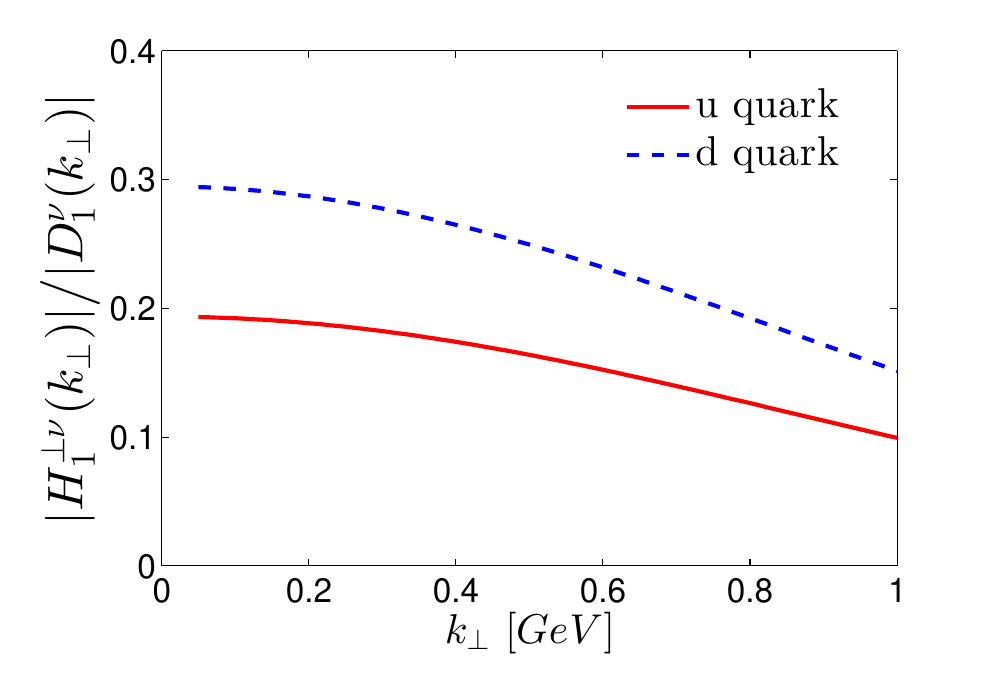}
\caption{\label{fig_R_FF} Ratio of the integrated(over $z$) Collins function 
and the integrated unpolarized fragmentation 
functions(Eq.\ref{FF_D1},\ref{FF_H1}) for $u$ and $d$ quarks.}
\end{figure}

\section{Contribution of $uu$ diquark}
The role of $ss$ diquarks was recently emphasized \cite{Karliner:2017kfm} in the studies of heavy baryons spectroscopy.  It is therefore instructive to explore the role of diquark containing light identical quarks.
To do so, we compared the results with and without (putting $C_{VV}=0$) $uu$ 
diquarks. Although the results do not change significantly, some disagreement 
for $z$ dependence of Collins asymmetry for $\pi^-$ mesons can be 
observed(Fig.\ref{fig_R_Pretz}). 

\begin{figure}[htbp]
\includegraphics[width=15.2cm,clip]{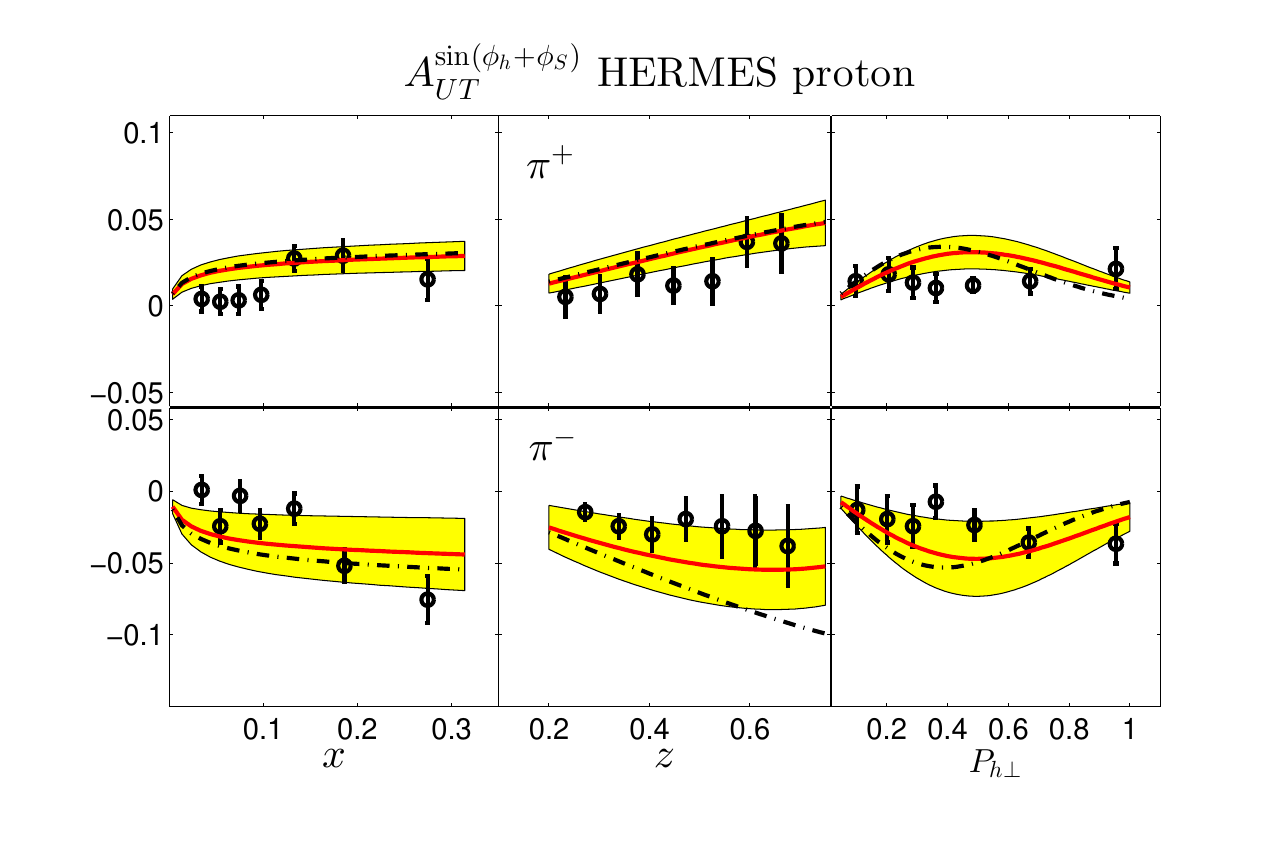}
\vspace{-1cm}
\caption{\label{fig_R_Pretz} Collins asymmetries of proton with and without 
$uu$ diquark. The black dot-dashed lines represent the asymmetry without  
contribution from $uu$ diquark($C_{VV}=0$).}
\end{figure}


\section{Summary and conclusion}
Azimuthal spin asymmetries are very important to understand  the three 
dimensional structure of the proton. There are many experimental as well as
theoretical model predictions for these asymmetries. 
Here, we have presented the results for both single and double spin asymmetries 
associated with T-even TMDs in a light front 
quark-diquark model of the proton for SIDIS processes in both $\pi^+$ and 
$\pi^-$ channels.  The results are compared with 
COMPASS and HERMES data. Since the experimental data are available at different energy scales, the scale dependence of the asymmetries which comes 
through the 
scale evolutions of the TMDs and FFs are required to be properly incorporated.
But unfortunately, the scale evolutions of all the TMDs are not yet 
 known. 
The scale evolutions of the polarized TMDs are not well understood, and in an approximation, they are assumed to be the same as unpolarized TMDs.
We have considered the scale evolution
of the unpolarized TMD only for which the QCD evolution is known to compare our results with experimental data at different energy scales. So, that 
the uncertainties in our results are 
confined within the  polarized TMDs only.
We have also compared  the results with
different approximate  evolution schemes for polarized TMDs.
There are four single spin asymmetries generated by Collins function but 
 $A^{\sin(\phi_h)}_{UL}$ and $A^{\sin(2 \phi_h)}_{UL}$ are generated by the 
same TMD and FF and hence are not treated independently. Our model 
predictions for  Collin asymmetry 
$A^{\sin(\phi_h+\phi_S)}_{UT}$ show good agreement with both HERMES and COMPASS  
data. The   amplitude of the 
asymmetry $A^{\sin(3\phi_h-\phi_S)}_{UT}$ 
 in our model is found to be suppressed by powers of $P_{h\perp}/M$ 
compared to the other SSAs and expected to be very small. Experimental data 
also show that the  average amplitude of the 
asymmetry $A^{\sin(3\phi_h-\phi_S)}_{UT}$ to be approximately zero.
 We have also predicted an 
asymmetry for the future EIC experiment. Our model predicts sizable Collins 
asymmetry $A_{UT}^{\sin(\phi_h-\phi_S)}$ for both $\pi^+$  and $\pi^-$
channels for the EIC experiments.

The double spin asymmetry $A_1^p$ in DIS depends on the PDFs  rather than  TMDs. Our model predictions for $A_1^p$ show excellent 
agreement with the data.  When the lepton beam is longitudinally polarized but the proton is transversely polarized, both  $A_{LT}^{\cos(\phi_h-\phi_S)}$ 
and   $A_{LT}^{\cos(\phi_h-\phi_S)}$ in the model are consistent with the 
experimental data and are found to be almost zero. But, the DSA when both proton 
and lepton beams are longitudinally polarized, $A_{LL}$ is quite large for both 
$\pi^+$ and $\pi^-$ channels which is also predicted in our model.

 We have explored different relations among the SSAs and DSAs and found an 
inequality similar to Soffer bound for PDFs. It will be interesting to see 
if similar relations are also found in other models.

\acknowledgements{ T.M thanks BLTP, Dubna for warm hospitality and organizers 
of the Helmholtz International Summer School "Hadron Structure and Hadronic 
Matter and Lattice QCD" 
 for the financial support to attend the summer school during 
which a part of the work has been done.}

\end{document}